\documentclass[11pt]{article}
\pdfoutput=1
\usepackage{graphicx,rotating,slashed,amsmath,xcolor,amssymb,amsfonts,colortbl,cite,bbold}
\usepackage{jcappub}
\usepackage{mathrsfs}
\usepackage{xcolor}
\usepackage{caption}
\usepackage{subcaption}
\usepackage{physics}

\newcommand{\kk}{{\bf k}}

\newcommand{\es}{\epsilon_s}

\def\gsim{ \lower .75ex \hbox{$\sim$} \llap{\raise .27ex \hbox{$>$}} }
\def\lsim{ \lower .75ex \hbox{$\sim$} \llap{\raise .27ex \hbox{$<$}} }
\def\be{\begin{equation}}
\def\ee{\end{equation}}
\def\bea{\begin{eqnarray}}
\def\eea{\end{eqnarray}}

\newcommand{\ba}{\begin{array}}
	\newcommand{\ea}{\end{array}}

\newcommand{\commentout}[1]{}

\newcommand{\comment}[1]{}

\newcommand{\bs}{\begin{split}}

	\def\ba{\begin{eqnarray}}
	\def\ea{\end{eqnarray}}
	\def\nn{\nonumber}

	\def\({\left(}
	\def\){\right)}

	\definecolor{jn}{RGB}{10, 10, 200} 
	\definecolor{js}{RGB}{204, 0, 0} 
	\definecolor{pgf}{RGB}{10, 150, 10} 
	\newcommand*{\mathcolor}{}
	\def\mathcolor#1#{\mathcoloraux{#1}}
	\newcommand*{\mathcoloraux}[3]{%
		\protect\leavevmode
		\begingroup
		\color#1{#2}#3%
		\endgroup
	}
	\newlength{\stheight}
	\newcommand\textst[1][fu-grey]{
		\ifmmode\setlength{\stheight}{+1.0ex}
		\else\setlength{\stheight}{+0.5ex}
		\fi
		\bgroup\markoverwith{\textcolor{#1}{\rule[\the\stheight]{2pt}{1.0pt}}}\ULon
	} 
	\newcommand{\textins}[2][fu-grey]{
		\ifmmode\mathcolor{#1}{#2}
		\else\textcolor{#1}{#2}\@\,
		\fi
	}
	\graphicspath{{./}}
	\allowdisplaybreaks
	
\providecommand{\abs}[1]{\lvert#1\rvert} 
\newcommand{\vect}[1]{\mathbf{#1}} 

\begin{document}
	\title{Non-Gaussian Signatures of a Thermal Big Bang}
	
	\author[a]{Maria Mylova}
	\author[b]{, Marianthi Moschou}
	\author[c,d,e]{, Niayesh Afshordi}
	\author[f]{and Jo\~ao Magueijo}

    \affiliation[a] {Cosmology, Ewha Womans University, 52 Ewhayeodae-gil, Seoul, Republic of Korea}
    \affiliation[b] {Department of Mathematics, Manchester University, Manchester, UK}
	\affiliation[c]{Department of Physics and Astronomy, University of Waterloo, Waterloo, ON, N2L 3G1, Canada}
	\affiliation[d]{Waterloo Centre for Astrophysics, University of Waterloo, Waterloo, ON, N2L 3G1, Canada}
	\affiliation[e]{Perimeter Institute for Theoretical Physics, 31 Caroline St. N., Waterloo, ON, N2L 2Y5, Canada}
	\affiliation[f]{Theoretical Physics Group, Blackett Laboratory, Imperial College, London, SW7 2BZ, UK}

	\emailAdd{nafshordi@pitp.ca}
	\emailAdd{j.magueijo@imperial.ac.uk}
	\emailAdd{marianthi.moschou@manchester.ac.uk}
	\emailAdd{mmylova@ewha.ac.kr}

	\date{today}
	\abstract{What if Big Bang was hot from its very inception? This is possible in a bimetric theory where 
		the source of fluctuations is thermal, requiring the model to live on a critical boundary in the space of parameters and can be realized when an anti-DBI brane moves within an  $EAdS_2 \times E_3$ geometry. 
 This setup renders the model unique, with sharp predictions for the scalar spectral index and its running. 	We investigate the non-Gaussian signatures of this thermal bimetric model, or ``bi-thermal'' for short. 
		We adapt the standard calculation of non-Gaussianities for $P(X,\phi)$ models to the thermal nature of the model, emphasising how the bi-thermal peculiarities affect the calculation and alter results.
		This leads to precise predictions for the shape and amplitude of the three-point function of the bi-thermal model
		(at tree-level): $f^{\rm local} _{\rm NL} = -3/2$ and $f^{\rm equil} _{\rm NL} = -2 + 4 \sqrt{3}\pi/9 \simeq 0.4$. We also discover a new shape of flattened non-gaussianity $\propto (k_1+k_2-k_3)^{-3/2} +$ permutations, which is expected due to the excited thermal initial conditions. These results, along with our earlier predictions for the scalar power spectrum, provide sharp targets for the future generation of cosmological surveys.  

	}
	
	\maketitle
	
	
	\section{Introduction}

	The inflationary paradigm as proposed in \cite{PhysRevD.23.347} correctly accounts for the Cosmic Microwave Background (CMB) anisotropies as observed by Planck \cite{Planck2020}. The  initial condition is provided by quantum fluctuations which freeze during a rapid phase of the exponential expansion of the universe. This provides the seeds responsible for the structure formation of the universe as we know it today.  
	
	An alternative approach is to consider that the initial conditions were thermal in nature, as it was originally proposed in \cite{JM2003}. In this case the universe starts in a thermal bath of finite temperature $T$ (see also~\cite{PedrThermal,BranTherma}). Assuming the thermal fluctuations experience a varying speed of sound \cite{JM2008,WK2010,NA2014}, where the acoustic horizon ($k c_s < H a$) rapidly shrinks, modes that originally were oscillating find themselves frozen, i.e., outside of the Hubble horizon ($k< aH$).  
		\begin{figure}[h]
		\includegraphics[width=\textwidth]{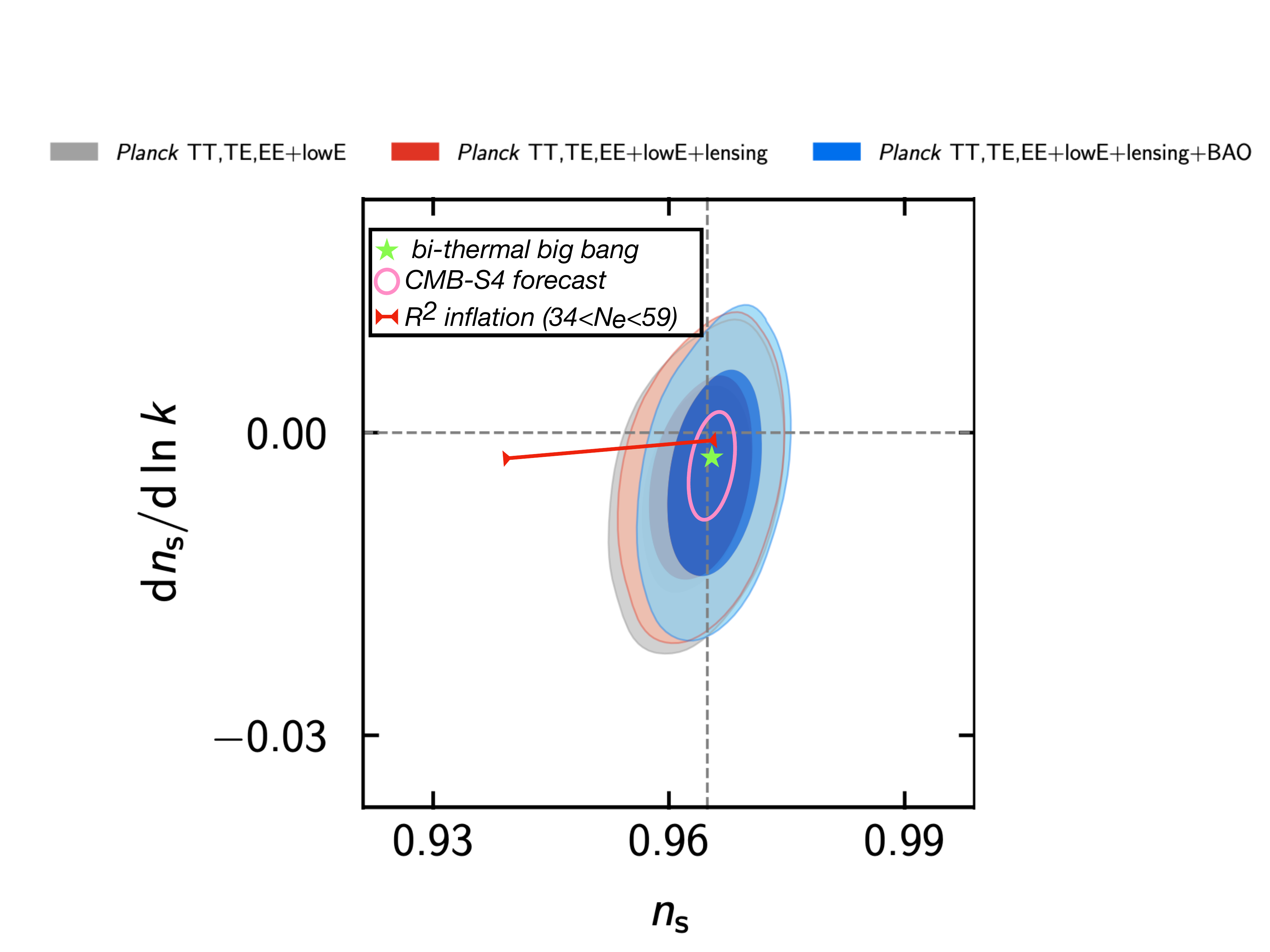}
		\caption{Current (forecasted) constraints on spectral index and running from Planck 2018++ \cite{Planck2020} (CMB S4 \cite{S4}), compared to predictions from bi-thermal model (Equation \ref{predict}) and $R^2$ (Starobinsky) inflation \cite{1980PhLB...91...99S}, given the allowed reheating range.}\label{Planck_fig}
	\end{figure}
	
	In this work, we focus on the thermal anti-DBI model, ``bi-thermal''for short, previously discussed in \cite{Critical}. The bi-thermal model is very interesting in that all its free parameters and predictions 
	are fundamentally fixed \cite{Critical}:
	\be
	n_s =  0.96478(64), ~~
\frac{dn_s}{d\ln k}
= -1.8 \times 10^{-3},\label{predict}
	\ee
	by the 
	amplitude of the scalar fluctuations, $A_s \simeq 2.1 \times 10^{-9}$ (at 0.05 /Mpc) \cite{Planck:2018jri} (see Fig. \ref{Planck_fig}).
	This is partly because the fluctuations in this scenario 
	have a thermal origin, and thus the model dispensing with a reheating phase, in contrast to inflationary models. In the latter,
	the reheating temperature (or number of e-foldings) introduces a constrained uncertainty in the predictions, for example,
	of the scalar spectral index $n_s$ and gravitational waves~\cite{macarena}. 
	
	Obviously, inflationary models could in principle have all the properties of reheating fixed by knowledge of the 
	particle physics parameters controlling the reheating process. However, these parameters are typically
	extraneous to the field producing inflation, so that this claim can only be substantiated in the rare cases where
	inflation is embedded in 
	a given particle physics model, such as the Higgs inflation model~\cite{2011JHEP...01..016B}. None of these problems affects
	the predictivity of the bi-thermal model.

	By mapping it into an ``Einstein frame'', the bi-thermal model can be understood as an anti-DBI model~\cite{JM2008,Bimetric,BimetricNG}. But as much as the model differs from inflation, it is also to be contrasted with these alternative theories in that 
	the model is ``critical'', in the sense that it lives on a discontinuous boundary. The models in~\cite{Bimetric,BimetricNG} 
	(based on a vacuum quantum fluctuations) do not have
	this property, so that deviations from exact scale-invariance can be continuously dialled~\cite{BimetricNG}. In the critical thermal model not only is {\it exact} scale invariance non-achievable, but the discontinuity imposes a very specific pattern of deviations from scale-invariance, with the amplitude of the fluctuations fixing where, on the steep slope of the discontinuity, the fluctuations we do observe
	must lie. This fixes the spectral index $n_s$ of the observed fluctuations, now deemed to be not only ``close to 1'', or ``generically red'', but also with an $n_s$ and its scale-dependence, fixed to several figures (see equation (\ref{predict})). The 
	combination of the two predictions renders the model a perfect target for upcoming experiments and observations~\cite{S4}. 
	
	One important matter that was not investigated in~\cite{Critical} is the non-Gaussian signature of these models. This aspect is
	particularly important for the {\it non-critical}, vacuum based anti-DBI models, because it is one of the few aspects in terms
	of which they are predictive~\cite{BimetricNG}. Working out the 3-point function for the critical model of~\cite{Critical} is the
	purpose of this paper. 
	
	To start with, the matter looks trivial. Famous last words... The problem is remarkably complex and this is the reason for the ungainly time delay between the publication of~\cite{Critical} and the present paper. The trivial aspects of the problem are presented in Sections~(\ref{bithmodel}) and (\ref{vertices}), where we review the two-point function set up and its extension to the vertices of the theory.  The 3-point function is evaluated in Section (\ref{3pfunc}). Unfortunately, the down to earth aspects of the calculation that follows the laying out of the formalism are far from trivial, as we find in Sections (\ref{bispec}) and (\ref{shapes}). This does not happen for particularly deep reasons, but (at the technical level) boils down to the fact that we need to evaluate integrals of triple products of Bessel functions away from the orders usually found in other models. Hence, a number of assumptions behind the approximations made in the literature are simply not valid here. 
	
	Nonetheless, using a combination of analytical and numerical methods we are able to bring the calculation to good port. The remarkable conclusions are as follows: the  non-Gaussianity of the bi-thermal model is described by three non-Gaussianity shapes; equilateral, squeezed and a new flattened shape. Therefore, the bi-thermal model can leave several non-Gaussian fingerprints in the early universe, with the bispectrum in the equilateral limit having an amplitude $f_{NL}^{\text{equil}} =   -   \frac{ 2}{9}  \qty(9 + 2 \sqrt{3}\pi)$, while in the squeezed limit  $f_{NL}^{\rm local} = - \frac{3}{2}$, both well in agreement with current observations \cite{Planck:2019kim}. The flattened non-Gaussianity is found to be $\propto (k_1+k_2-k_3)^{-3/2}$. Finally, we conclude in Section (\ref{conclu}). A technical Appendix can be found in Section (\ref{App}).

	\section{The bi-thermal model}\label{bithmodel}
	
	\subsection{The model}
	
	The bi-thermal model, as presented in \cite{Critical}, can be nicely written as a scalar-tensor theory that takes on the following `anti-DBI' form
	\be \label{action}
	S = \int d^4x \sqrt{-g}\left[\frac{1}{B(\phi)}\sqrt{1+2B(\phi)X}-V(\phi)\right],
	\ee
	where $\phi$ is a scalar field, $X \equiv -\partial_\mu \phi \partial^\mu \phi$ is its kinetic term (here we are using a mostly + metric convention) and $g$ is the metric determinant. The `anti-DBI' name tag derives from the fact that this is essentially a DBI action with an opposite sign warp factor $B(\phi)$. 
	Using Bianchi identities to fix the form of the potential and fixing the residual ambiguity by choosing a so-called ``critical solution'' (for details see \cite{Critical}) we arrive at a particular form for the warp factor and potential
	\begin{align}
	\nn B(\phi) &= B_0 \left(\frac{\phi}{M_{\rm Pl}}\right)^2 \\
	V(\phi) &= \frac{3}{4 B_0} \ln \left(\frac{\phi}{M_{\rm Pl}}\right)^2.
	\label{pot}
	\end{align}
	With this as model-specific input, one can then use standard results for $P(X,\phi)$ theories to investigate the correlation functions this model gives rise to in the early universe.
	
	When considering the non-Gaussian signatures associated with the bi-thermal model, three features of this model and its `critical' solution \eqref{pot} are worth highlighting at this point. 
	Firstly, the model can be understood as coming from the DBI action of a 3-brane embedded in a $EAdS_2 \times E_3$ geometry. As such the action will be protected by a non-linear realisation of the associated symmetries and the particular DBI-like form will in fact lead to one of the vertices in the cubic action vanishing -- we will discuss this in Section (\ref{vertices}). Secondly, no scaling solutions, so no exactly scale invariant solutions can be found, which is a special feature of the potential \eqref{pot}.\footnote{Other power-law warp factors lead to power-law potentials with scaling solutions.}  This is important, because analytic expressions for the non-Gaussian amplitude of a $P(X,\phi)$ theory are easiest to derive for exact scale invariance \cite{KP}, in which case slow-roll need not be assumed, or up to leading order in slow-roll parameters \cite{ChenNG}, in which case scale-invariance need not be assumed. Finally, the speed of sound of such models will tend to infinity/diverge in the UV/at early times. As such it will also be varying rapidly and a slow-roll approximation becomes inappropriate (for closely related non-Gaussian computations in more standard $P(X,\phi)$ models, see \cite{BimetricNG,FastRoll}). We will therefore be interested in a particular setup, where we have an almost, but not exactly scale-invariant solution and an extremely large speed of sound at the time CMB correlations are sourced. 
	

\subsection{The quadratic action and mode functions}

With an eye on computing non-Gaussian statistics later on (i.e. three-point statistics descending from the cubic effective action in perturbations), it is useful to quickly consider the quadratic regime first and identify the mode functions that will also enter higher-point statistics. We define the following two variables
\begin{align}
q &\equiv \frac{a \sqrt{2\epsilon}}{\sqrt{c_s}} ,
&y  &\equiv \int \frac{c_s dt}{a} = \int c_s d\eta,  \label{qy_def}
\end{align}
where $q$ is expressed in terms of the scale factor $a$, the speed of sound of scalar perturbations $c_s$ and the standard slow-roll parameter $\epsilon  \equiv  - \dot{H}/H^2$, while the  `sound horizon time' variable $y$ is related to physical time $t$ and conformal time $\eta$. In terms of these we can define the so-called Mukhanov variable (a canonically-normalized scalar variable) $v = M_{\rm Pl} q\zeta$, where $\zeta$ is the usual coming curvature perturbation. This way we can write the Mukhanov-Sasaki equation
\begin{equation}
v''_k + \left( k^2 - \frac{q''}{q} \right) v_k = 0,
\end{equation}
which governs the evolution of $\zeta$ at the level of the quadratic action and where the $q$ parameter, we already encountered above, satisfies
\be
\frac{q''}{q} = \frac{\nu^2 -1/4}{y^2}. \label{q-Q}
\ee
As such, the mode equation becomes
\begin{equation}
v''_k + \left[ k^2 - \frac{\nu^2 -1/4}{y^2} \right] v_k = 0. \label{mode_beta}
\end{equation}
In the usual Bunch-Davies solution, one would set $\nu = 3/2$ in order to recover scale invariance. However, for the thermal case we are considering here, we have $\nu = 1$ for a scale-invariant solution. This means we now have a modified differential equation for $q$, with solutions
\be
q \sim (-y)^{-1/2}  \qquad {\rm or}\qquad q \sim y^{3/2} \,,
\label{qsiT}
\ee
where $y$ runs from $-\infty$ to $0$, so the growing mode solution is now $q \sim (-y)^{-1/2}$. Correspondingly, we now have the following solution for the Mukhanov variables
\be
v_{k}(y) = \frac{\sqrt{-\pi y}}{2} e^{-i\gamma_\nu}H^{(2)}_\nu (ky),\label{hankel}
\ee
where $\gamma_\nu = \frac{\pi}{4}(2\nu+1)$ and $\nu \sim 1$ for near scale-invariant solutions, as discussed above. 

The transition from a ``v'' variable to a ``u'' variable (directly linked to $\zeta$) is now given by~\cite{KP,BimetricNG}:
\bea \label{uexp}
u_k(y) &=& \frac{1}{a}\left( \frac{c_s}{2\epsilon}\right)^{1/2} v_k(y)\nn\\
&=& \frac{1}{a}\left( \frac{c_s}{2\epsilon}\right)^{1/2} \frac{\sqrt{-\pi y}}{2}e^{-i\gamma_\nu} H^{(2)}_\nu (ky),
\eea
where $\gamma_\nu = \frac{\pi}{4}(2\nu+1)$ and where the argument $\nu$ of the Hankel function will eventually be set to (approximately) one, implying a (near) scale-invariant two-point function. 
At a constant {\it comoving} temperature $T_c = a T/c_s$, the power spectrum of scalar perturbations is given by:
\be
{\cal P}_{\zeta}(k)= \lim_{y \rightarrow 0^-} \frac{k^3}{2 {\pi}^{2}} \frac{{\left| v_k \right|}^2}{q^2M^2_P} [2 \langle n_k\rangle_{T_c} + 1] \simeq \frac{c_s T_c (-y)^{1-2\nu}}{2\pi^3  a^2 \epsilon M^2_P}\bigg|_{y\sim -k^{-1}},\label{power}
\ee
where we used the Bose-Einstein distribution for phonon occupation number $\langle n_k\rangle_{T_c} \simeq T_c/k$ for $k \ll T_c$.  

\section{Non-Gaussian vertices}\label{vertices}

Having computed the mode functions above, we are now in a position to start working with the cubic effective action and to compute non-Gaussian signals. 

\subsection{Preliminaries}

We would like to evaluate the three-point function in the interaction picture, following the procedure outlined by \cite{Maldacena,ChenNG,SeeryLidsey}. In the interaction picture we are, therefore, after
\begin{equation} \label{interaction}
\langle
\zeta(t,\textbf{k}_1)\zeta(t,\textbf{k}_2)\zeta(t,\textbf{k}_3)\rangle=
-i\int_{t_0}^{t}{\rm d}t^{\prime}\langle[
\zeta(t,\textbf{k}_1)\zeta(t,\textbf{k}_2)\zeta(t,\textbf{k}_3),H_{\rm int}(t^{\prime})]\rangle ~,
\end{equation}
where we will specify the range of integration later. This result is true to first order in $H_{\rm int}$ (higher order corrections can be computed using `nested' commutators with $H_{\rm int}$) and for the cubic interactions we will be interested in $H_{\rm int} = -{\cal L}_{\rm int}$, after all terms proportional to the equations of motion have been removed via field redefinitions.

In order to explicitly compute \eqref{interaction}, we need to expand the curvature perturbation $\zeta$ in terms of raising and lowering operators, as usual
\begin{equation} \label{modes} 
\zeta(y, \kk) = u_k(y)a(\kk) + u_k^*(y) a^\dagger(-\kk).
\end{equation} 
Note that the mode functions $u_k$ in \eqref{modes} satisfy the corresponding Mukhanov-Sasaki equation and we will be applying the commutation relations $[a(\kk), a^\dagger(\kk')] = (2 \pi)^3 \delta^3(\kk - \kk')$, where the factor of $(2\pi)^3$ implicitly fixes our Fourier convention. 

\subsection{Leading order vertices}

The cubic effective action, from which we will calculate the non-Gaussian amplitudes, is the usual \cite{Maldacena,ChenNG,SeeryLidsey}
\begin{eqnarray} \label{action3}
S_{(3)}&=& M_{\rm Pl}^2 \int {\rm d}t {\rm d}^3x \left\{
-a^3 \left[\Sigma\left(1-\frac{1}{c_s^2}\right)+2\lambda\right] \frac{\dot{\zeta}^3}{H^3}
+\frac{a^3\epsilon}{c_s^4}(\epsilon-3+3c_s^2)\zeta\dot{\zeta}^2 \right.
\nonumber \\ &+&
\frac{a\epsilon}{c_s^2}(\epsilon-2\es+1-c_s^2)\zeta(\partial\zeta)^2-
2a \frac{\epsilon}{c_s^2}\dot{\zeta}(\partial
\zeta)(\partial \chi) \nonumber \\ &+& \left.
\frac{\epsilon}{2a}(\partial\zeta)(\partial
\chi) \partial^2 \chi +\frac{\epsilon}{4a}(\partial^2\zeta)(\partial
\chi)^2+ 2 f(\zeta)\left.\frac{\delta L_{(2)}}{\delta \zeta}\right\vert_1 \right\} ~.\nn \\
\end{eqnarray}
Now, the first vertex contribution $(\dot \zeta^3)$ vanishes for anti-DBI (and indeed DBI) models and the final term can be removed via a field redefinition. Here, dots denote derivatives with respect to proper time $t$, $\partial$ is a spatial derivative, we have explicitly imposed that $\eta = \dot{\epsilon}/H\epsilon = 0$ and $\chi$ is defined via
\begin{equation}
\partial^2 \chi = \frac{a^2 \epsilon}{c_s^2}\dot{\zeta}\,.
\end{equation}
We now perform the field re-definition removing the final term (this only introduces additional terms proportional to $\eta$, i.e. terms which we will drop), so that for our model we are left with
\begin{eqnarray} \label{action31}
S_{(3)}&=& M_{\rm Pl}^2 \int {\rm d}t {\rm d}^3x \left\{
\frac{a^3\epsilon}{c_s^4}(\epsilon-3+3c_s^2)\zeta\dot{\zeta}^2 \right. +
\frac{a\epsilon}{c_s^2}(\epsilon-2\es+1-c_s^2)\zeta(\partial\zeta)^2-
2a \frac{\epsilon}{c_s^2}\dot{\zeta}(\partial
\zeta)(\partial \chi) \nonumber \\ &+& \left.
\frac{\epsilon}{2a}(\partial\zeta)(\partial
\chi) \partial^2 \chi +\frac{\epsilon}{4a}(\partial^2\zeta)(\partial
\chi)^2\right\} ~,
\end{eqnarray}
We can now re-write this, explicitly, in terms of $y$-time and find the following cubic effective action
\begin{eqnarray} \label{action32}
S_{(3)}&=& M_{\rm Pl}^2 \int {\rm d}y {\rm d}^3x \frac{a}{c_s}\left\{
\frac{a\epsilon}{c_s^2}(\epsilon-3+3c_s^2)\zeta{\zeta}^{'}{}^2 \right. +
\frac{a\epsilon}{c_s^2}(\epsilon-2\es+1-c_s^2)\zeta(\partial\zeta)^2 \nonumber \\ &-&
2\frac{\epsilon}{c_s}{\zeta}^{'}(\partial
\zeta)(\partial \hat\chi) + \left.
\frac{\epsilon}{2a}(\partial\zeta)(\partial
\hat\chi) \partial^2 \hat\chi +\frac{\epsilon}{4a}(\partial^2\zeta)(\partial
\hat\chi)^2\right\} ~,
\end{eqnarray}
where ${\zeta}^{'} \equiv (d/dy)\zeta$ and we now have 
\begin{equation}
\partial^2 \hat\chi = \frac{a \epsilon}{c_s}{\zeta}^{'}\,.
\end{equation}
Ignoring the leading $1/c_s$ factor in \eqref{action32}, all terms in the second line go as $1/c_s^2$, whereas the top line has contributions independent of $c_s^2$. In the large $c_s$ limit, which is what we are interested in here, these terms will dominate. So we can write
\begin{eqnarray} \label{action33}
S_{(3)}^\text{leading order}&=& M_{\rm Pl}^2 \int {\rm d}y {\rm d}^3x \frac{a^2 \epsilon}{c_s}\left\{3\zeta{\zeta}^{'}{}^2  -\zeta(\partial\zeta)^2 \right\}.
\end{eqnarray}
Note that this means the non-Gaussian amplitude is `slow-roll suppressed' in this model, although of course there is no requirement that $\epsilon \ll 1$ here, so this is not necessarily a suppression.

\section{Direct Evaluation of 3-point function}
\label{3pfunc}

In this section, we will directly compute the 3-point function, starting from the third order action. Let us combine the quadratic action for $\zeta$ with the leading correction (\ref{action33}):
\be
S = \frac{M_{\rm Pl}^2 }{2} \int dy  d^3x q^2(y) \left[ (1+3 \zeta)\zeta'^2 - (1+\zeta) (\partial \zeta)^2  + {\cal O}(\zeta^4) \right]. 
\ee
The Euler-Lagrange equations are then given by:
\be
q^{-2}[q^2(1+3\zeta)\zeta']' = \nabla \cdot [(1+\zeta)\nabla\zeta] + 3\zeta'^2-(\partial\zeta)^2 +{\cal O}(\zeta^3),
\ee
which simplifies to
\be
q^{-2}(q^2\zeta')' -\partial^2 \zeta = -3\zeta\left[q^{-2}(q^2\zeta')' -\partial^2 \zeta\right]  - 2\zeta \partial^2\zeta +{\cal O}(\zeta^3).
\ee
Using the first order equation of motion, the first term on the right hand side vanishes at second order, which yields:
\be
q^{-2}(q^2\zeta')' -\partial^2 \zeta = - 2\zeta \partial^2\zeta +{\cal O}(\zeta^3),
\ee
and equivalently in Fourier space:
\be
q^{-2}(q^2\zeta_{\bf k}')' +k^2 \zeta_{\bf k} = 2\int \frac{d^3 k'}{(2\pi)^3} |{\bf k - k'}|^2 \zeta_{\bf k'}  \zeta_{\bf k- k'}  +{\cal O}(\zeta^3).
\ee
The next step is to solve this equation perturbatively. The retarded Green's function for the linear $\zeta$ equation can be written in terms of the mode function $u_{k}(y)$
\begin{equation}\begin{split} 
G_k(y,w) &= \frac{q^2(w)\left[u_k(y)u_k^*(w)-u_k^*(y)u_k(w)\right]}{q^2(y)\left[u'_k(y)u_k^*(y)-u'^*_k(y)u_k(y)\right]} \Theta(y-w) 
\\&=  i  q^2(w)\left[u_k(y)u_k^*(w)-u_k^*(y)u_k(w)\right] \Theta(y-w),
\label{eq:3pa} 
\end{split}\end{equation}
%
which can be used to solve the second order equation of motion, at leading order:
\begin{equation}\begin{split} 
\zeta^{(2)}_{\bf k}(y) &= \zeta^{(1)}_{\bf k}(y) + 2  i \int_{-\infty}^y dw q^2(w) \int  \frac{d^3 k'}{(2\pi)^3} |{\bf k - k'}|^2 
\\& \times \left[u_k(y)u_k^*(w)-u_k^*(y)u_k(w)\right]\zeta^{(1)}_{\bf k'}(w)  \zeta^{(1)}_{\bf k- k'}(w).
\label{eq:3pb} 
\end{split}\end{equation}
%
%
We can now compute the 3-point function:
\begin{equation}\begin{split} 
\lim_{y\rightarrow 0^-} \langle  \zeta_{\bf k_1} \zeta_{\bf k_2} \zeta_{\bf k_3} \rangle = (2\pi)^3  \delta^3({\bf k_1+k_2+k_3})  (5/3)^3 B_{\Phi}(k_1,k_2,k_3)
\label{eq:3p} 
\end{split}\end{equation}
where the bispectrum is given by
\begin{equation}\begin{split} 
 B_{\Phi}(k_1,k_2,k_3) &=-\frac{432 \pi^4 P^2_{\zeta}}{125 (k_1 k_2 k_3)^2} 
 \Bigg\{\int_{-\infty}^0 dw \, q^2(w)   k^2_1(k_2^2+k_3^2)
\\&  \times \frac{\Im\left[u_{k_1}(0)u_{k_1}^*(w)\right]  \Re\left[u_{k_2}(0) u_{k_2}^*(w)\right] \Re\left[u_{k_3}(0) u_{k_3}^*(w)\right] }{k_2 k_3 |u_{k_2}(0)|^2 |u_{k_3}(0)|^2} + {\rm 2~ perm.} \Bigg\}.
\label{eq:bis} 
\end{split}\end{equation}
%

\section{Evaluating the bispectrum} \label{bispec}

For now, we focus on the integral 
\begin{equation}\begin{split} 
\mathcal{I} &= \int^0_{-\infty}  \dd{w} q^2(w) \frac{\Im[u_{k_1}(0) u^*_{k_1}(w)] \Re[u_{k_2}(0) u^*_{k_2}(w)]\Re[u_{k_3}(0) u^*_{k_3}(w)]}{\abs{u_{k_2}(0)}^2 \abs{u_{k_3}(0)}^2},
\label{eq:i6} 
\end{split}\end{equation}
where we have omitted the permutations and  $i\epsilon$ prescription. Taking the real and imaginary parts gives
\begin{equation}\begin{split} 
\mathcal{I} &= -i \int^0_{-\infty}  \dd{w} q^2(w) 
 \frac{[u_{k_1}(0) u_{k_1}^*(w)-u_{k_1}^*(0)u_{k_1}(w)] [u_{k_2}(0) u_{k_2}^*(w)+u_{k_2}^*(0)u_{k_2}(w)] }{8  u_{k_2}(0) u_{k_2}^*(0) u_{k_3}(0) u_{k_3}^*(0)}
\\&\times [u_{k_3}(0) u_{k_3}^*(w)+u_{k_3}^*(0)u_{k_3}(w)]. 
\label{eq:i3} 
\end{split}\end{equation}
%
%
%
We change the domain of integration to $(0, \infty)$ by making explicit that in conformal time $w=-\abs{w}$ (from here onwards we drop the absolute sign). We use the analytic continuation formula (10.11.8) in \cite{abramowitz+stegun}, given by
\begin{equation}\begin{split} 
H_n^{(2)}(z e^{m\pi i}) = (-1)^{mn} \qty(m H_n^{(1)}(z) + (m+1) H_n^{(2)}(z)),
\label{eq:i1} 
\end{split}\end{equation}
which allows us to take the complex conjugates
\begin{equation}\begin{split} 
H_\nu^{(1)}(\bar{z}) = \overline{H_\nu^{(2)}(z)}, \qq{and} H_\nu^{(2)}(\bar{z}) = \overline{H_\nu^{(1)}(z)},
\label{eq:i2} 
\end{split}\end{equation}
for positive argument and real order $\nu$. The mode functions become
\begin{equation}\begin{split} 
& u_k(w) = -\frac{1}{q(w)} \frac{\sqrt{\pi w}}{2} e^{\displaystyle - \frac{i 3 \pi}{4}} \qty(H^{(1)}_\nu(k w)+2 H^{(2)}_\nu(k w)),
\\& u_k^*(w) = - \frac{1}{q(w)} \frac{\sqrt{\pi w}}{2} e^{\displaystyle  \frac{i 3 \pi}{4}} \qty(H^{(2)}_\nu(k w)+2 H^{(1)}_\nu(k w)),
\label{eq:i4} 
\end{split}\end{equation}
which results to
\begin{equation}\begin{split} 
\lim_{w\rightarrow 0} u_k(0)= \frac{(-1)^{\frac{3}{4}}}{k \sqrt{\pi}},
\label{eq:i5} 
\end{split}\end{equation}
after the small argument limit has been considered. 
Then, using expressions \eqref{eq:i1}--\eqref{eq:i5}, one can simplify the integral \eqref{eq:i3} to 
\begin{equation}\begin{split} 
\mathcal{I} &= - \frac{3 k_2 k_3 \pi^2}{64 k_1} \int_0^{\infty}  \dd{w}  w^2 \qty(H^{(1)}_\nu(k_1 w)+H^{(2)}_\nu(k_1 w)) 
\\& \times \qty(H^{(1)}_\nu(k_2 w)-H^{(2)}_\nu(k_2 w)) \qty(H^{(1)}_\nu(k_3 w)-H^{(2)}_\nu(k_3 w)) 
\\& = \frac{3 k_2 k_3 \pi^2}{8 k_1} \int_0^{\infty}  \dd{w}  w^2  J_1(k_1 w) Y_1(k_2 w) Y_1(k_3 w). 
\label{eq:i8}
\end{split}\end{equation}
In the last step, we set the order of the Hankel functions to $\nu=1$ and expanded in terms of Bessel functions and simplified. This resulted to an integral over a triple product of Bessel functions. Thus,
the 3-point function is now written as\footnote{From here onwards we follow the conventions in \cite{2016} and drop the $4 \pi^2$ factor from the bispectrum expression.}
%
\begin{equation}\begin{split} 
B_{\Phi}(k_1,k_2,k_3) &= -  (2 \pi)^3 \delta^3(\vect{k}_1 + \vect{k}_2 + \vect{k}_3 ) \qty(\frac{5}{3})^3 \frac{1}{(k_1 k_2 k_3)^3} \mathcal{P}_\zeta^2 \frac{108}{125} k_1^3 (k_2^2+k_3^2) 
 \\& \times \frac{3 k_2 k_3 \pi^2}{8 k_1} \int_0^{\infty}  \dd{w}  w^2  J_1(k_1 w) Y_1(k_2 w) Y_1(k_3 w)  +  \qq{2 perms}.
\label{eq:i7} 
\end{split}\end{equation}
%
The integrals involved in our analysis are highly oscillatory and divergent,
causing a great complexity in the analytical and numerical methods of integration. Therefore, alternative approaches should be used to deal with this issue. Next, we examine each shape individually, i.e. we independently address the equilateral, local and isosceles configurations.

\subsection{The equilateral configuration}

To obtain the equilateral shape we let $k_1=k_2=k_3=k$. Then, the integral in \eqref{eq:i8} simplifies to 
\begin{equation}\begin{split} 
\mathcal{I}_{equil} &=  \frac{3 k \pi^2}{8} \int_0^{\infty}  \dd{w}  w^2  J_1(k w) Y_1^2(k w) ,
\label{eq:i9} 
\end{split}\end{equation}
which can be evaluated exactly, namely $\mathcal{I}_{equil} = \frac{9 + 2 \sqrt{3}\pi}{18 k^2}$.
Finally, accounting for all permutations, together with the prefactor involved in (\ref{eq:i7}), the bispectrum in the equilateral configuration is given by the expression
\begin{equation}\begin{split} 
B_\Phi &=  - (2 \pi)^3 \delta^3(\vect{k}_1 + \vect{k}_2 + \vect{k}_3 ) \frac{\mathcal{P}_\zeta^2}{(k_1 k_2 k_3)^3}  \frac{4}{3}  k^3 \qty(9+2 \sqrt{3} \pi).
\label{eq:i9b} 
\end{split}\end{equation}
We can also verify the above result by  solving the integral, numerically, mode by mode. To this end, we firstly need to regulate the oscillatory nature of \eqref{eq:i9}, at infinity, by multiplying the integrand with $e^{- \epsilon \abs{w}^2/2}$, where $\epsilon \ll 1$. This has the effect of suppressing the highly oscillatory behaviour at infinity and forcing convergence of the numerical integration (see Appendix (\ref{Ap2})). We find that for very small values of $\epsilon$, the numerical results are in full agreement with the analytic results for a large range of comoving scales which could be observed with current and future CMB experiments. 

As we will see later, the equilateral configuration is subdominant since the bispectrum peaks in the squeezed and folded configurations. Nevertheless, discussing it in detail allows us to introduce the regulating procedure, due to the simplicity of the problem, in this case. Next, we will use this method in place of the $i \epsilon$ prescription, to demonstrate convergence of the analytic result, for the bispectrum amplitude, in the squeezed limit. 

\subsection{Local configuration}

Next, we move on to the local configuration by letting the two long wave modes to be equal, that is $k_1=k_2 =k$ and take the limit of the remaining mode to zero, $k_3 \rightarrow 0$. 
Similar to what presented in the previous subsection, there are three permutations of momenta to consider, two of which coincide owing to the symmetry of the triple product of the Bessel functions involved in the original integrand \eqref{eq:i8}. Therefore, after taking the small argument approximation of the corresponding Bessel function for $k_3\to 0$, the two integrals to be solved are
\begin{equation}\begin{split} 
&  \mathcal{I}_{sq_1} = -\frac{3  \pi}{4} \int_0^{\infty}  \dd{w}  w  J_1(k w) Y_1(k w),
\label{eq:i11} 
\end{split}\end{equation}
and 
\begin{equation}\begin{split} 
&  \mathcal{I}_{sq_2} = \frac{3}{16} \pi^2 k^2  \int_0^{\infty}  \dd{w}  w^3 Y_1^2(k w). 
\label{eq:i11a} 
\end{split}\end{equation}
First, let us focus on the integral \eqref{eq:i11}. Using the following formula  (see Section 10.22. in \cite{abramowitz+stegun})
\begin{equation}\begin{split} 
\int \dd{z} z \mathscr{B}_\mu (a z)  \mathscr{D}_\mu (a z)  = \frac{1}{4} z^2 \qty[2 \mathscr{B}_\mu (a z)  \mathscr{D}_\mu (a z) - \mathscr{B}_{\mu+1} (a z)  \mathscr{D}_{\mu+1} (a z)-\mathscr{B}_{\mu+1} (a z)  \mathscr{D}_{\mu-1} (a z)   ],
\label{eq:i12} 
\end{split}\end{equation}
allows us to compute the integral \eqref{eq:i11} in terms of its antiderivative. Thus,
\begin{equation}\begin{split} 
&  \mathcal{I}_{sq_1} =  \frac{3  \pi}{4} \left( \lim_{w\to\infty} \mathcal{I}_{sq1}^{\text{anti}}(w) - \mathcal{I}_{sq1}^{\text{anti}}(0) \right),
\label{eq:i13} 
\end{split}\end{equation}
where
\begin{equation}\begin{split} 
\mathcal{I}_{sq_1}^{\text{anti}}(w)  =\frac{1}{4} w^2 \qty[J_2( k w) Y_0(k w) -2 J_1(k w) Y_1(k w) + J_0(k w) Y_2(k w)].
\label{eq:i13} 
\end{split}\end{equation}
Finally, the approximations for large and small arguments of Bessel functions can be used accordingly, such that 
\begin{equation}
    \mathcal{I}_{sq_1} = \frac{3 }{4 k^2}  - \frac{3 }{8 k^2} \sin(2 k w)  + \mathcal{O}(w),
\end{equation}
where the second term, involving the sine function, is the contribution from infinity. 

As expected, the last result recovers the oscillations at infinity, which should be regulated after applying the $i\epsilon$ prescription, leaving behind the leading order term, that is
\begin{equation}
    \mathcal{I}_{sq_1} = \frac{3 }{4 k^2} +\mathcal{O}(w),\quad\text{ as }w\to 0.
    \label{eq:i14}
\end{equation}
Alternatively, the oscillatory terms at infinity can be regulated by computing the result numerically, following the approach discussed in the equilateral configuration case. That is,  to multiply the integrand in \eqref{eq:i11} by a \textit{regulator} of the form $e^{- \epsilon \abs{w}^2/2}$, $\epsilon \ll 1$, which tames the highly oscillatory terms and forces convergence (see Appendix (\ref{Ap2})). Indeed, our numerical result is in full agreement with the analytic result in (\ref{eq:i14}).

Finally, including the prefactor and accounting for both permutations, gives
\begin{equation}\begin{split} 
 B_{\phi_{sq_1}} &=  - (2 \pi)^3 \delta^3(\vect{k}_1 + \vect{k}_2 + \vect{k}_3 ) \frac{1}{(k_1 k_2 k_3)^3} \mathcal{P}_\zeta^2 6 k^3.
\label{eq:i16a} 
\end{split}\end{equation}
Next, we consider the remaining permutation, $  \mathcal{I}_{sq_2}$. Using the formula (see section 10.22. in \cite{abramowitz+stegun})
\begin{equation}\begin{split} 
& \int \dd{z} z^{\mu+\nu+1} \mathscr{B}_\mu (a z)  \mathscr{D}_\nu (a z)  =  \frac{z^{\mu+\nu+2}}{2(\mu+\nu+1)} \qty[\mathscr{B}_\mu (a z)  \mathscr{D}_\nu (a z)+ \mathscr{B}_{\mu+1} (a z)  \mathscr{D}_{\nu+1} (a z)], 
\label{eq:i12a} 
\end{split}\end{equation}
where $\mu+\nu \neq -1$, we find the antiderivative
\begin{equation}\begin{split} 
\mathcal{I}^{\text{anti}}_{sq_2} = \frac{3}{16} \pi^2 k^2 \times \frac{1}{6} w^4 \qty(Y^2_1(k w) + Y^2_2(k w)).
\label{eq:i17} 
\end{split}\end{equation}
Likewise to our analysis above, the appropriate approximations of the Bessel functions  can be used, along with a regulation process, resulting to 
\begin{equation}\begin{split} 
\mathcal{I}_{sq_2} =  \frac{1 }{2 k^2} + \mathcal{O}(w),\quad\text{ as }w\to 0.
\label{eq:i18} 
\end{split}\end{equation}
Putting it together with the prefactor (it is customary to ignore the $(k_1 k_2 k_3)^{-3}$ term in the bispectrum amplitude calculations when taking the squeezed limit, $k_3 \rightarrow 0$) gives a vanishing result in the limit $k_3\rightarrow0$ 
\begin{equation}\begin{split} 
&\lim_{k_3 \rightarrow 0}  B_{\phi_{sq_2}} \simeq \frac{108}{125} k_3^3 \bigg|_{\lim k_3 \rightarrow 0} = 0.
\label{eq:i21a} 
\end{split}\end{equation}
In other words, the result that contributes to the bispectrum amplitude, in the squeezed limit, is the one obtained from the first two permutations in (\ref{eq:i16a}). 

Note that  again we found  the main contribution to the bispectrum comes from approximating the integral near the lower limit. This can be also understood in another way. The subhorizon modes are oscillating and cancel out, leaving only the dominant contribution for $0< w \lesssim w_\Lambda $ where $w_\Lambda$ is some cut-off. 
In general, the cutoff $w_\Lambda \sim 1/k$ gives results that are consistent with analytic calculations, but as we shall see in Section (\ref{shapes}), this fails in the flattened limit where there is resonance between the modes (i.e. oscillations don't cancel) and thus a physical cutoff will be needed to regulate the integrals. 

 

\subsection{Isosceles configuration}

The last case to be discussed is the isosceles configuration. In principle, this is the most general case, from which one can derive all the other standard shapes. Therefore, solving for the isosceles configuration will not only give a more general result, but also verify all the previous findings, i.e those for the equilateral and local configurations.

By conservation of momentum (ensured by the argument of the $\delta$-function) we have three permutations of the momenta of the closed triangle $k_1,k_2,k_3$. As before, two of them coincide due to the symmetry of the triple product of the Bessel functions involved in the original integrand \eqref{eq:i8}. That is, setting $k_1=k_2 = k$, the integrals to solve are

%
\begin{equation}\begin{split} 
\mathcal{I}_{iso_1} &=  \frac{3  k_3 \pi^2}{8 } \int_0^{\infty}  \dd{w}  w^2  J_1(k w) Y_1(k w) Y_1(k_3 w),  
\label{eq:i8a} 
\end{split}\end{equation}
and
\begin{equation}\begin{split} 
\mathcal{I}_{iso_2} &=  \frac{3 k^2 \pi^2}{8 k_3} \int_0^{\infty}  \dd{w}  w^2  J_1(k_3 w) Y_1^2(k w), 
\label{eq:i8bb} 
\end{split}\end{equation}
where the integral \eqref{eq:i8a} is repeated twice when all three permutations are summed.
The integrals \eqref{eq:i8a}--\eqref{eq:i8bb} can be expressed in terms of the very general Meijer G-functions (see Appendix in (\ref{Ap1})) and then be evaluated as
\begin{equation}\begin{split} 
\mathcal{I}_{iso_1} &=
-\frac{3 \pi^{3/2}}{2 k_3^2} \text{MeijerG} \left[\Big\{ \{ 0  \},\{1/2,1/2 \}\Big\} ,\Big\{ \{ 0,1 \},\{1/2 \}\Big\}, \frac{4k^2}{k_3^2}
\right]
\\& = \frac{3 }{2 k_3} \times \left[ \left( \frac{1}{
 4k^2 - k_3^2} \right)
  \left( k3 + \frac{4k \text{ arccsc}\left(\frac{2k}{k_3}\right)}{\sqrt{4-\frac{k3^2}{k^2}}} \right)
  \right],
\label{eq:i8b} 
\end{split}\end{equation}
and
\begin{equation}\begin{split} 
\mathcal{I}_{iso_2} &=    \frac{3  k^2 \pi^{3/2}}{k_3^4} \text{MeijerG} \left[\Big\{ \{ -1  \},\{-1/2,1/2 \}\Big\} ,\Big\{ \{ -1,1 \},\{-1/2 \}\Big\}, \frac{4k^2}{k_1^2}
\right]
\\& = \frac{6 k^2 k_3 -3 k_3^3 - \frac{24 k^3 \text{arccsc}\left(\frac{2k}{k_3}\right)}{\sqrt{4-\frac{k_3^2}{k^2}}}}{8 k^2 k_3^3 - 2 k_3^5},
\label{eq:i8b1} 
\end{split}\end{equation}
%
%
where both \eqref{eq:i8b} and \eqref{eq:i8b1} are valid for $k_3<2k$. Finally, the total result for the bispectrum in the isosceles limit, including the correct prefactors for each permutation, is
\begin{equation}\begin{split} 
B_{\Phi_{iso}} &= - (2 \pi)^3 \delta^{(3)}(\vect{k}_1+\vect{k}_2+\vect{k}_3) P_\zeta^2  \frac{1}{(k_1 k_2 k_3)^3} \frac{1}{k_3 \qty(4 k^2 - k_3^2)^\frac{3}{2}}
\\& \times \qty[12 k^2 \qty(k_3 \sqrt{4 k^2 - k_3^2} \qty(k^3 +2 k^2 k_3 + k k_3^2 - k_3^3)+4 k^3 (k-k_3)^2 \arccsc\qty(\frac{2k}{k_3}))] 
\label{eq:itot} 
\end{split}\end{equation}
%
%
%
 From the last expression we can derive the squeezed limit by letting $k_3\to 0$, as well as any other shapes in between. In particular, we will find the bi-thermal bispectrum peaks in the flattened limit, that is the limit when $k_3\to 2k$. Indeed, one can verify that by taking $k_3 \rightarrow 0$ in (\ref{eq:itot}), we obtain the local configuration expression
\begin{equation}\begin{split} 
B_{\Phi_{sq}} &=  -  (2\pi)^3  \delta^{(3)}(\vect{k}_1+\vect{k}_2+\vect{k}_3) \frac{1}{(k_1 k_2 k_3)^3} P_\zeta^2 6 k^3,
\label{eq:isq1} 
\end{split}\end{equation}
which fully agrees with our result in (\ref{eq:i16a}). Note that adopting this approach, which involves Meijer G-functions and their general definition via a line integral in the complex plane (see Appendix (\ref{Ap1})) does not necessitate the use of a regulation process at infinity, since this is dealt with the complex analysis arguments. Also, we find that the second permutation, given in (\ref{eq:i8b1}), does not contribute to the bispectrum amplitude in this limit, which is in agreement with our result in (\ref{eq:i21a}).

\section{Bispectrum results}
\label{shapes}

\begin{figure}
	\begin{subfigure}[b]{0.45\textwidth}
		\centering
		\includegraphics[width=\linewidth]{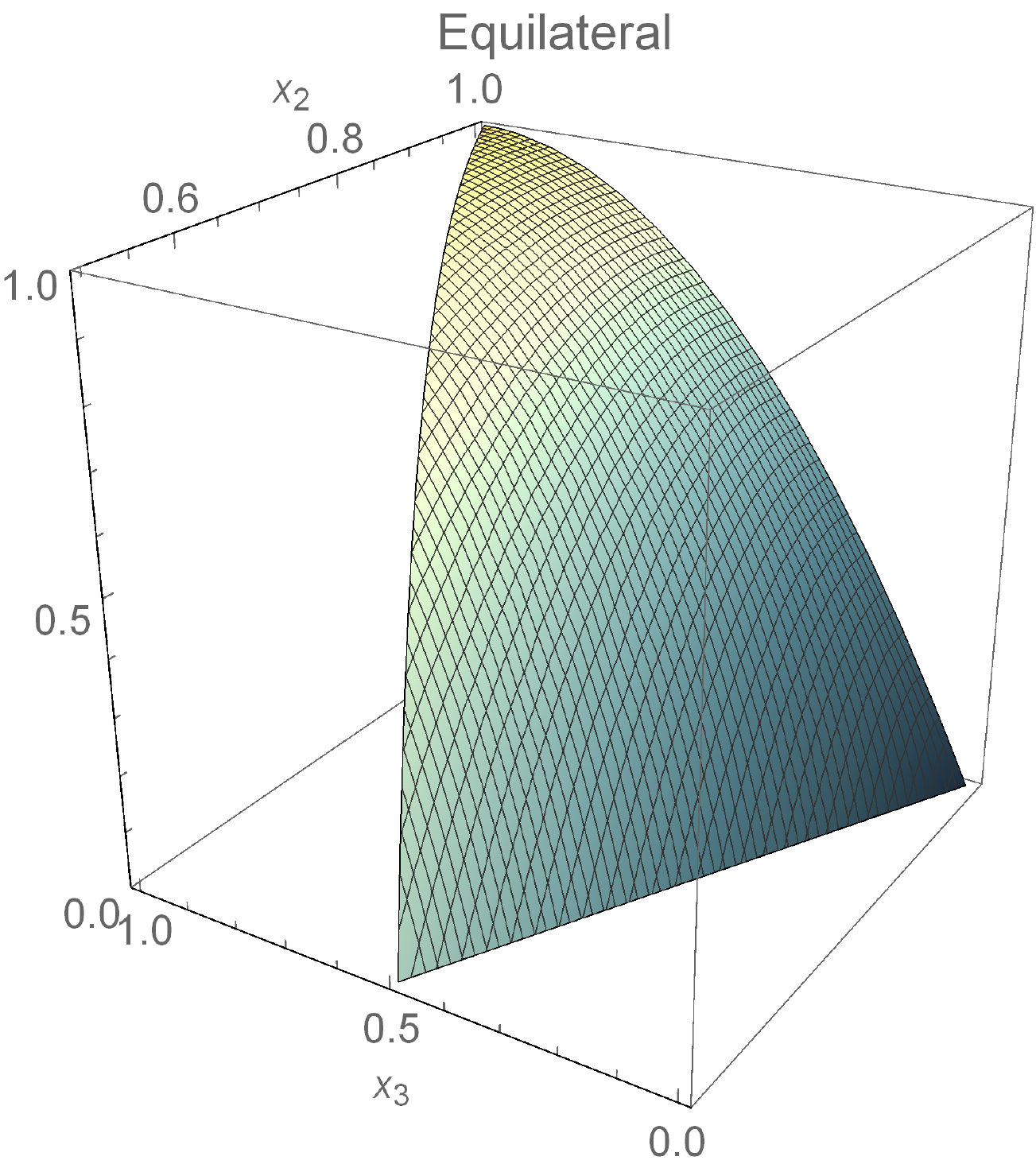}
		\caption{ Equilateral}
		\label{fig:eq1}
	\end{subfigure}\hfill
	\begin{subfigure}[b]{0.45\textwidth}
	\centering
	\includegraphics[width=\linewidth]{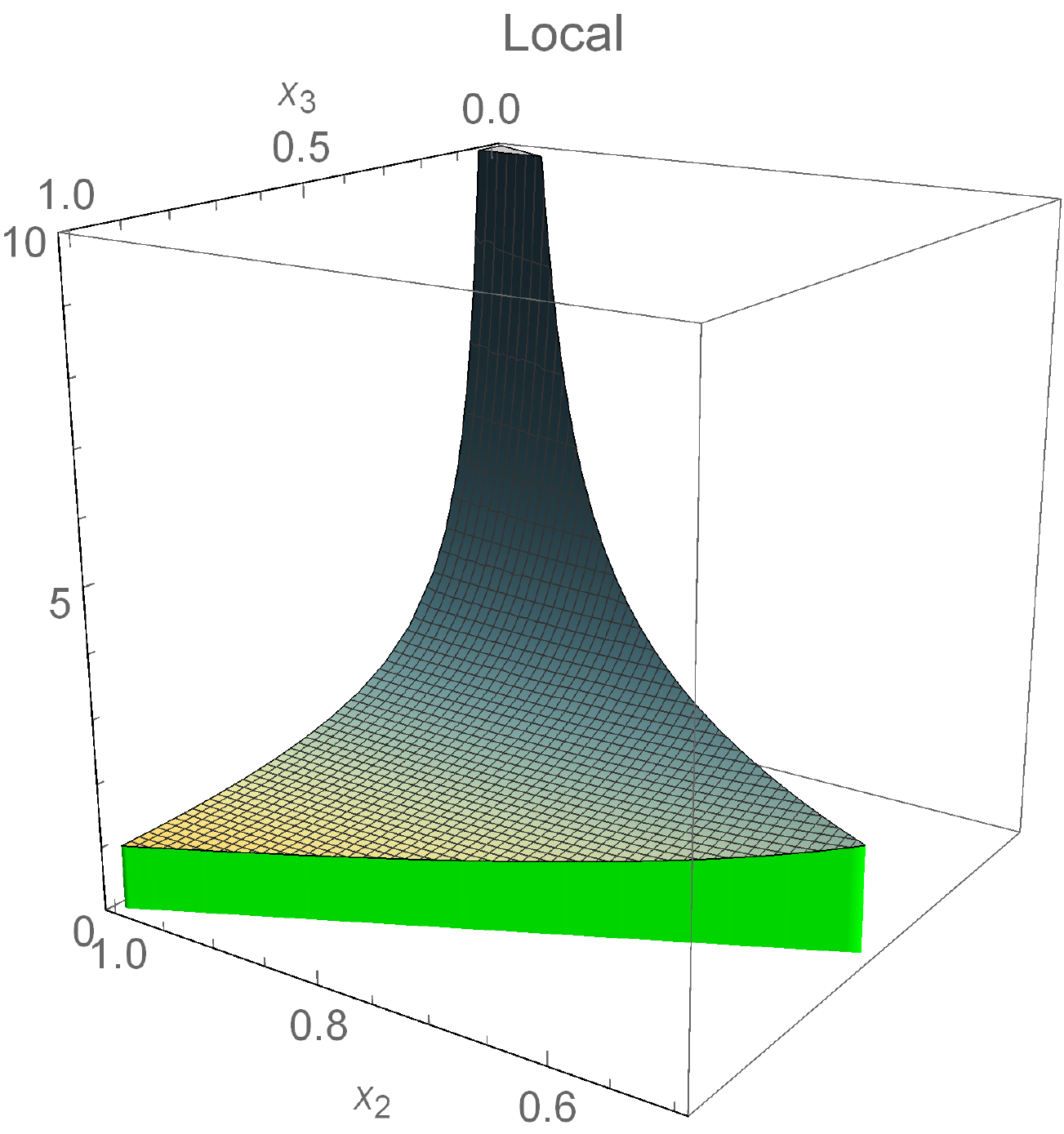}
	\caption{Local}
	\label{fig:logc}
\end{subfigure}
	\caption{Plot of the shape function in the equilateral (equation \ref{eq:i10c}) and local (Equation \ref{eq:ng8}) configuration (normalised). } \label{fig:local1}
\end{figure}

Let us first recall the following formulae for calculating the amplitude and shape of the bispectrum \cite{2016}. The bispectrum is defined, as
\begin{equation}\begin{split} 
\langle \Phi(\vect{k}_1) \Phi(\vect{k}_2) \Phi(\vect{k}_3)\rangle = (2\pi)^3 \delta^{(3)}(\vect{k}_1+\vect{k}_1+\vect{k}_1) B_\Phi(k_1,k_2,k_3),
\label{eq:ng1}
\end{split}\end{equation}
where $\Phi = \frac{3}{5}\zeta$. Assuming scale invariant statistics, the shape function is defined, as 
\begin{equation}\begin{split} 
S(k_1,k_2,k_3) =  \frac{(k_1 k_2 k_3)^2}{A^2} B_\Phi(k_1,k_2,k_3),
\label{eq:ng1a}
\end{split}\end{equation}
where $A$ is the dimensionless power spectrum. 
We first look at the equilateral limit. Using our result in (\ref{eq:i9b}), we see that in the equilateral limit the bispectrum goes like $B_\phi \sim k^{-6}$. Therefore, we can use the standard template 
\begin{equation}\begin{split} 
B_\Phi(k_1,k_2,k_3) &= 6 A^2 f_{NL}^{\text{equil}} \bigg[-\qty(\frac{1}{k_1^3 k_2^3}+\frac{1}{k_2^3 k_3^3}+\frac{1}{k_3^3 k_1^3}) - \frac{2}{(k_1 k_2 k_3)^2} + \bigg(\frac{1}{k_1 k_2^2 k_3^3} + \frac{1}{k_1 k_3^2 k_2^3} 
\\& +\frac{1}{k_2 k_1^2 k_3^3} + \frac{1}{k_2 k_3^2 k_1^3} +\frac{1}{k_3 k_1^2 k_2^3}+ \frac{1}{k_3 k_2^2 k_1^3} \bigg)\bigg].
\label{eq:ng2}
\end{split}\end{equation}
 Setting $k_1 = k_2 = k_3=k$ in the above, gives $B_\Phi(k_1,k_2,k_3) = 6 A^2 f_{NL}^{\text{equil}} k^{-6}$. Identifying $A$ with $P_\zeta$, we find the bispectrum amplitude in the equilateral limit, is given by \footnote{The negative sign in the bispectrum amplitudes simply indicates negative correlation of the fluctuations.}
\begin{equation}\begin{split} 
f_{NL}^{\text{equil}} &=   -   \frac{ 2}{9}  \qty(9 + 2 \sqrt{3}\pi). 
\label{eq:i10aa} 
\end{split}\end{equation}
Using (\ref{eq:ng1a}), the normalised shape in the equilateral configuration, takes the form
%
\begin{equation}\begin{split} 
\frac{S^\text{equil}(x_2,x_3)}{S^\text{equil}(1,1,1)} = \bigg[-2 + \frac{1}{x_2} + x_2 + \frac{1}{x_3} - \frac{1}{x_2 x_3} + \frac{x_2}{x_3} -\frac{x_2^2}{x_3} + x_3 + \frac{x_3}{x_2} - \frac{x_3^2}{x_2}\bigg],
\label{eq:i10c} 
\end{split}\end{equation}
where $x_2 = k_2 /k_1$ and $x_3 = k_3/k_1$. This is plotted in Fig. (\ref{fig:local1}). 

Next, using (\ref{eq:i16a}), we find that in the squeezed limit the bispectrum goes like $B_\Phi \sim k^3$. Therefore, we can use the standard template
	\begin{equation}\begin{split} 
B_\Phi^{\text{local}}(k_1, k_2, k_3) &= 2 A^2 f_{NL}^{\text{local}} \qty(\frac{1}{k_1^3 k_2^3} + \frac{1}{k_2^3 k_3^3} +\frac{1}{k_1^3 k_3^3})
\label{eq:ng7}.
\end{split}\end{equation}
Taking $k_3 \rightarrow 0, k_2 \sim k_1$ of the above, gives the bispectrum in the squeezed limit.
Comparing this to our result in (\ref{eq:i16a}), we find	
	\begin{equation}\begin{split} 
 f_{NL}^{\rm local} = - \frac{3}{2},
\label{eq:i33} 
\end{split}\end{equation}
which is well within experimental bounds (\cite{Planck:2019kim}).
Using (\ref{eq:ng1a}), the local shape is given by 
\begin{equation}\begin{split} 
S^{\text{local}}( x_2, x_3) & =  \frac{1}{3}f_{NL}^{\text{local}} \qty(\frac{1}{x_2 x_3}+ \frac{x_2^2}{x_3}+\frac{x_3^2}{x_2}),
\label{eq:ng8}
\end{split}\end{equation}
where we ensure this is normalised for $S(1,1,1,)=1$. This gives the familiar shape in Fig. (\ref{fig:local1}).


So far, we have found that in the equilateral and squeezed limits the bispectrum takes a standard form. This enabled us to obtain values for the bispectrum amplitude $f_{NL}$ in those limits. Next, we examine the general form of the expression in (\ref{eq:itot}). 

 Before we start, a few words are in order. Looking at the denominator in (\ref{eq:itot}), we find that there is a divergence at the limit $k_3 \to 2 k$. Now, if we naively restore the average of the $k$ momenta this becomes $k_3 \to k_1+k_2$, which is the flattened limit of the bispectrum. This effect is not entirely surprising. Such behaviour has been encountered before in \cite{2007}, when non-Bunch-Davies vacua were considered, as well in \cite{2008} they found that the bispectrum peaks in the flattened limit when starting from excited initial states (see also \cite{2010Fold,2021arXiv211006038A}). As we discussed above, the reason for this divergence is that excited modes, e.g., with acoustic frequencies $k_1$ and $k_2$ are in resonance with $k_3$, if $k_3 \simeq k_1+k_2$ and thus can excite $k_3$ particles in long before  horizon crossing. If the resonance is not perfect, we expect the number of $k_3$ particles excited to scale as an inverse power of $k_1+k_2-k_3$. Alternatively, the divergence can be limited by the finite time that the modes spend in the thermal bath. Indeed, the early thermal vacuum is understood as a collection of excited states in equilibrium while the sound horizon is taken to be infinite at $t \to 0$ (or $y \to -\infty$). If we want to be more realistic we may expect a hard cut-off or regulator, corresponding to a quantum gravity phase where the bi-thermal classical description is no longer valid. Solving numerically, this ensures a finite result in the folded limit for a wide range of modes  (in agreement with the conclusions in \cite{2007}).

 A possibly more surprising feature of our model is the appearance of local (or squeezed) non-Gaussianity, usually expected in multi-field models, even though we only have a single scalar field. However, we note the temperature of the field, effectively, acts as a second clock in this model, which allows for coupling of short and long wavelength modes.

\begin{figure}[htbp]
		\begin{center}
			\includegraphics[width=0.8\textwidth]{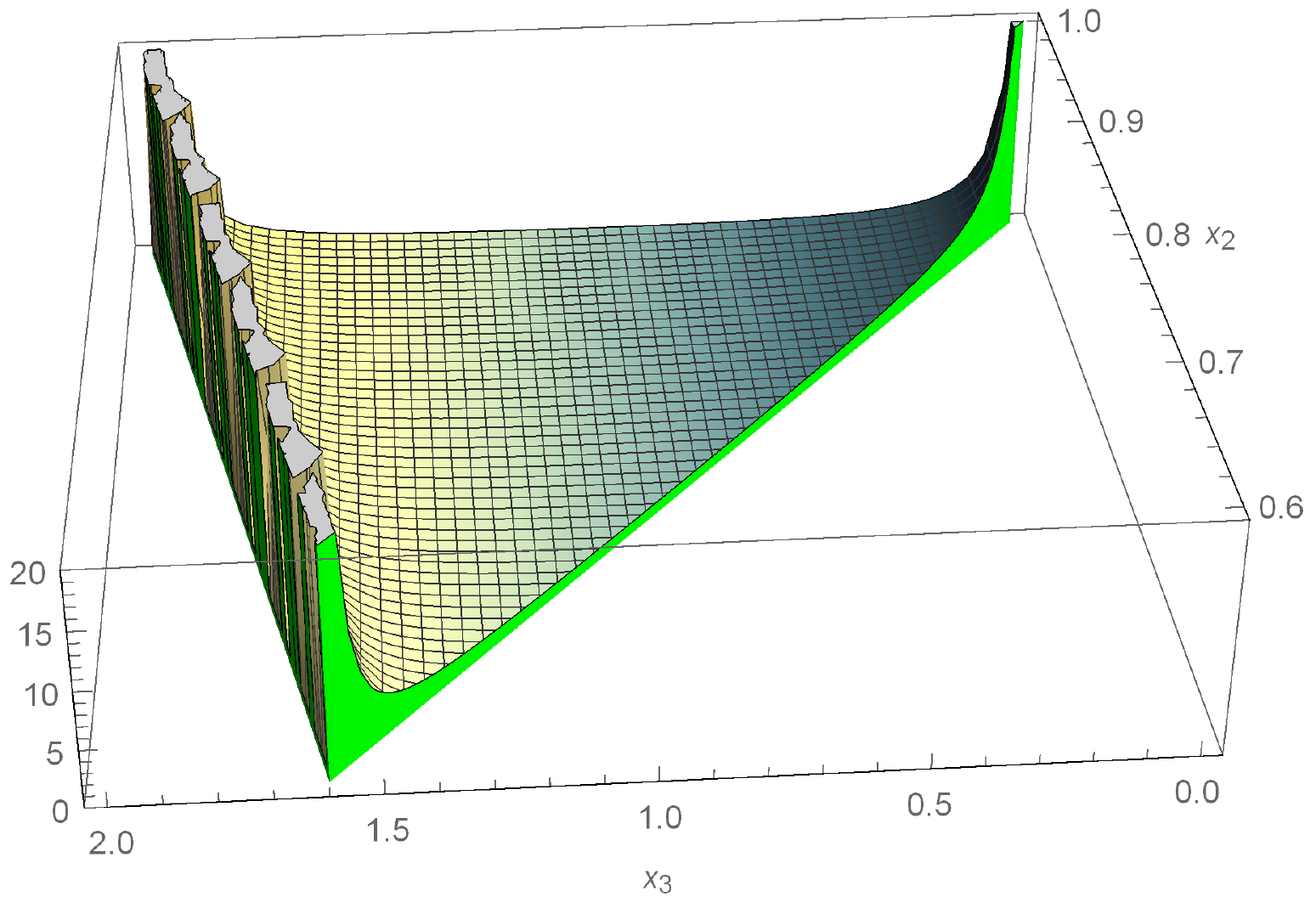}
			\caption{Our final bispectrum shape function for the bi-thermal model (equation \ref{eq:ng8}), extrapolated from exact isosceles result. The bispectrum peaks at the folded and squeezed limit.}
			\label{fig:fd}
		\end{center}
		\setlength{\abovecaptionskip}{0pt plus 1pt minus 1pt}
	\end{figure}

We proceed to examine the shape of (\ref{eq:itot}). We can naively restore the momenta as $k = 1/2 (k_1+k_2)$. Using this and substituting for $x_2= k_2/k_1$ and $x_3 = k_3/k_1$ in the standard expression for the bispectrum shape (\ref{eq:ng1a}), we obtain
\begin{equation}\begin{split} 
S_{\text{bi-thermal}}(x_2, x_3) & =- \frac{1}{8 x_2 x_3^2 \qty(1+x_2-x_3)^\frac{3}{2} \qty(1+x_2+x_3)^\frac{3}{2} } \Bigg[ 3 \qty(1+x_2)^2 
\\& \times \Bigg(x_3 \sqrt{\qty(1+x_2-x_3)\qty(1+x_2+x_3)} \bigg(\qty(1+x_2)^3+ 4 \qty(1+x_2)^2 x_3 
\\& + 4 \qty(1+x_2) x_3^2 - 8 x_3^3 \bigg) + \qty(1+x_2)^3 \qty(1+x_2-2x_3)^2 \arccsc\qty(\frac{1+x_2}{x_3}) \Bigg) \Bigg].
\label{eq:ng8}
\end{split}\end{equation}
%
%
The normalised shape is plotted in Fig. (\ref{fig:fd}). We see that the bispectrum peaks in the folded and squeezed limit, as expected. Please note, this is not a template for general $ x_2,x_3$ but only valid in the limit $x_3 \rightarrow 0$ and on the line $x_3 = 2x_2$.

Finally, setting $x_2=1$ and expanding around $x_3\sim2$, we find that in the flattened limit the bispectrum shape goes as $\propto (k_1+k_2-k_3)^{-3/2}$. 

	\section{Conclusions}
	\label{conclu}
	
	The bi-thermal model of~\cite{Critical} is the pinnacle of a proud lineage of models playing with the speed of light~\cite{VSLreview}, most notably rendering the speed of massless matter particles different from the speed of gravity. Starting from~\cite{Moffat0,AM,Barr} these models became progressively better defined~\cite{VSLcov} and more concrete and solid in their predictions~\cite{MoffatBi,JM2008,Bimetric}, eventually matching inflation. But it was not until~\cite{Critical} that this general class of models met with the prospect of potentially supplanting inflation in predictivity~\cite{macarena}. This claim rests on past work on the power spectrum of the fluctuations. To this we have now added predictions involving the 3-point function.

Specifically we found the following results:

\begin{itemize}
    \item Squeezed non-Gaussianity with $f^{\rm local}_{NL} = - \frac{3}{2}$, which is unique for single field scenarios.
    \item Equilateral non-Gaussianity with $f^{\rm equil}_{NL} =-   \frac{ 2}{9}  \qty(9 + 2 \sqrt{3}\pi)$.
    \item We obtained a new shape that diverges as $(k_1+k_2 - k_3)^{-\frac{3}{2}}$ in the flattened  limit.
\end{itemize}
These results are quite spectacular, in that they leave no leeway for fiddling with parameters, as they are  distinctive and unique to this model. Hence, not only is the bi-thermal model more predictive than inflation, but it is also capable of making a prediction which cannot be manufactured within inflation, without significant fine-tuning.

Although, it is true that current data can only, at most, disprove different classes of models, we hope that the unique qualities of the bi-thermal model combined with its rich non-Gaussian footprint will help, in the future, to differentiate it from other models of inflation. 
	
	

Finally, we would like to add that the presence of a divergence in the flattened limit necessitates the application of a physical cutoff. We argue that this divergence is simply an artefact of the effective field theory.  Indeed, there should be extra corrections coming from the effective field theory of the DBI action (or the geometric picture). It would be interesting to see how the flattened limit is modified in that case. Ideally, the UV completion of the bi-thermal model should, almost certainly, be able to regulate this divergence.

There are currently weak observational constraints for non-Gaussianities in the flattened limit. The observability of this signal is something interesting we should study in the future.

\section{Acknowledgments} We would like to thank Johannes Noller for initial collaboration and Jinn-Ouk Gong for discussions and advice in relation to this paper. This work was supported by 
the STFC Consolidated Grant ST/L00044X/1 (JM), and the National Research Foundation of Korea Grant 2019R1A2C2085023 (Maria M), the University of Waterloo, Natural Sciences and Engineering Research Council of Canada (NSERC) and the Perimeter Institute for Theoretical Physics (NA).
Research at Perimeter Institute is supported in part by the Government of Canada through
the Department of Innovation, Science and Economic Development Canada and by the
Province of Ontario through the Ministry of Colleges and Universities.

\appendix
		\section{Appendix}
		\label{App}
		
		\subsection{Analytic continuation of Meijer G-functions}
		\label{Ap1}
		
\begin{figure}
    \centering
    \includegraphics[scale=1]{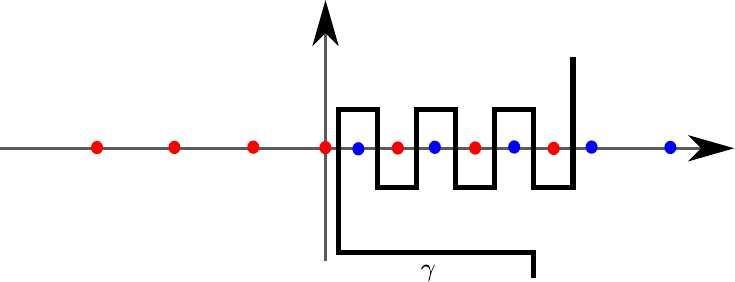}
    \caption{ The contour of a Meijer G-function lies between the poles of $\Gamma(1-a_i -s)$ and the poles of $\Gamma(b_i +s)$. }
    \label{fig1}
\end{figure}

		A Meijer G-function is defined as follows 
\begin{align}
        G_{p,q}^{m,n} \left( z \left| \,\begin{matrix}
a_1 & \cdots & a_p\\
b_1 & \cdots & b_q
\end{matrix} \right. \right) =&
\text{MeijerG}\left[\Big\{ \{ a_1,\dots a_n  \},\{a_{n+1},\dots a_p \}\Big\} ,\Big\{ \{ b_1,\dots b_m  \},\{b_{m+1},\dots b_q \}\Big\},z
\right] \nonumber
\\
=\frac{1}{2 \pi i}& \int_{\gamma} \frac{\Gamma(1-a_1 -s) \cdots \Gamma(1-a_n - s) \Gamma(b_1 + s) \cdots \Gamma(b_m +s)}{\Gamma(a_{n+1} +s) \cdots \Gamma(a_p + s) \Gamma(1- b_{m+1} - s) \cdots \Gamma(1 - b_q -s)}\, z^{-s} \text{d}s,
\label{generalmeijer}
\end{align}
where the contour $\gamma$ is illustrated in Fig.~(\ref{fig1}). 

In many special cases, Meijer G-function is converted to other known functions, as it can be seen from the evaluation of the Bessel integrals below. The contour integral can be viewed as an \emph{inverse Mellin transform} and can be treated as such. One can find the poles of the integrand and then by a \emph{residue analysis} can evaluate the integral as an infinite series. There should be paid attention to the domain of convergence, since depending on the values of $z$, only some of the poles will be considered in the integration process.

\begin{figure}
	\begin{subfigure}[b]{0.45\textwidth}
		\centering
		 \includegraphics[width=\linewidth]{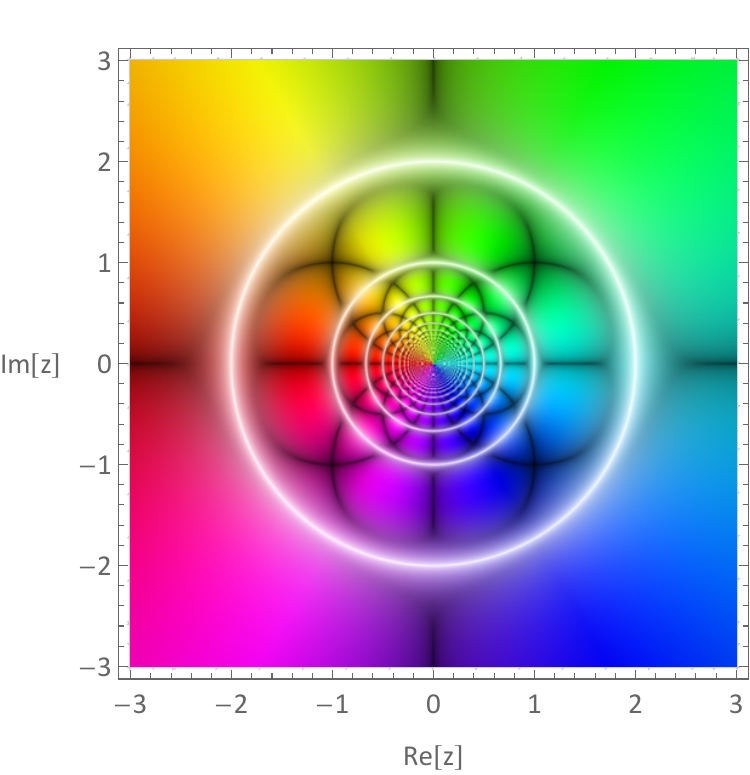}
		\caption{}
		\label{phaseplot1}
	\end{subfigure}\hspace{1em}
	\begin{subfigure}[b]{0.45\textwidth}
	\centering 
  \includegraphics[width=\linewidth]{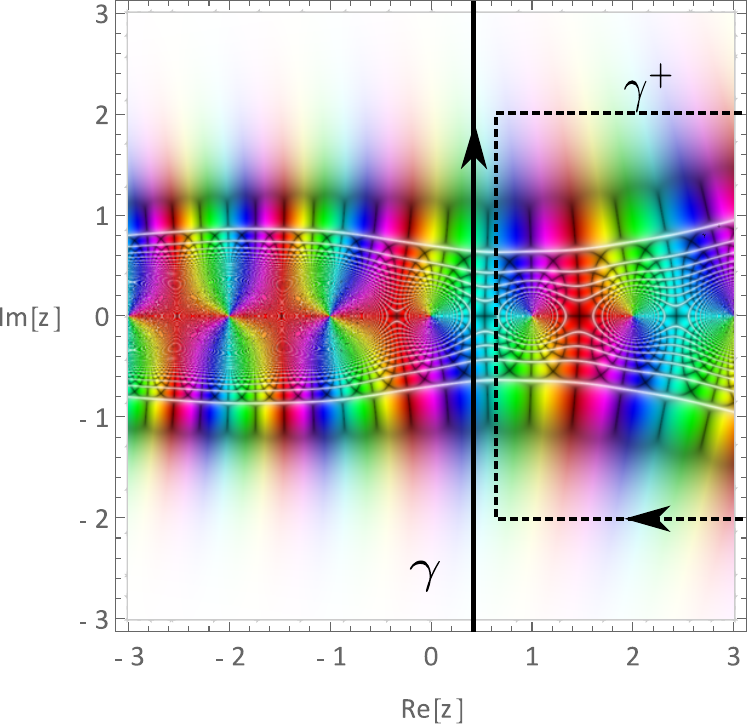}
	\caption{}
	\label{phaseplot2}
\end{subfigure}
	\caption{(a) An example of a phase-portrait. Here, we illustrate the phase-portrait of the complex function f(z) = 1/z (simple pole at z). The colour variation shows how the phase continuously varies while encircling a simple pole, i.e. red indicates negative real values ($\pm\pi$ phase), cyan the positive real values (zero phase), etc (b) A phase-portrait of the product of $\Gamma(1-s)\Gamma(s)\Gamma(1+s)  \left(4k^2/k_3^2\right)^{-s}$ can be seen (nominator in \eqref{withGamma1}). Here, $\frac{2k}{k_3}=1.5$, i.e.\  $2k > k_3$. The solid black line represents the original $\gamma$ contour and the dashed one the deformed contour. Note that in the negative real $z$-axis the non-convergence effect is clearly visible. Here, we used the complex variable $z$ instead of $s$.} \label{fig2}
\end{figure}

The integral in (\ref{eq:i8a}) evaluates to
\begin{align}
        \mathcal{I}_{iso_1}(k, k_3) =&
\frac{3  k_3 \pi^2}{8 } \int_0^{\infty} \text{d}w \,\,w^2 J_1 (k w) Y_1(k w)  Y_1(k_3 w)\nonumber\\
=& -\frac{3 \pi^{3/2}}{2 k_3^2} \text{MeijerG} \left[\Big\{ \{ 0  \},\{1/2,1/2 \}\Big\} ,\Big\{ \{ 0,1 \},\{1/2 \}\Big\}, \frac{4k^2}{k_3^2}
\right], \label{meijerrepresentation1}
\end{align}
which utilizes the Meijer G-function, as defined in \eqref{generalmeijer}, for $n=1$, $p=3$, $m=2$, $q=3$ and $z= \frac{4k^2}{k_3^2}$, with $a_1=a_n=0$, $a_2=a_{n+1}=1/2$, $a_3=a_p=1/2$ and  $b_1=0$, $b_2=b_{m}=1$, $b_3=b_{m+1}=b_q =1/2$. For the simplification from the integral form to the Meijer G representation we used \cite{adamchik1990algorithm}. Writing expression \eqref{meijerrepresentation1} in its line integral form, the original integral for $\mathcal{I}_{iso_1}$ is now written as
\begin{align}
        \mathcal{I}_{iso_1}(k, k_3) 
 =&-\frac{3 \pi^{3/2}}{2 k_3^2} \frac{1}{2\pi i} \int_{\gamma} \frac{\Gamma(1-s)\Gamma(s)\Gamma(1 +s)}{\Gamma(\frac{1}{2} +s) \Gamma(\frac{1}{2} + s) \Gamma(\frac{1}{2} - s))}\, \left(\frac{4k^2}{k_3^2}\right)^{-s} \text{d}s,\label{withGamma1}\\
 =& -\frac{3 \pi^{3/2}}{2 k_3^2} \frac{1}{2\pi i} \int_{\gamma} \frac{\cos(\pi s)}{\sin(\pi s)} \frac{\Gamma(1+s)}{\Gamma(\frac{1}{2} + s)}\left(\frac{4k^2}{k_3^2}\right)^{-s} \text{d}s,\label{contourintegral}
\end{align}
where the contour $\gamma$ can be seen in the left of Fig.~(\ref{fig2}). In the same plot, the poles of the nominator in  \eqref{withGamma1} can be seen. Note that for $2k = k_3$ the above integral does not converge. Therefore, we should proceed under the condition that either $2k < k_3$ or $2k > k_3$, such that convergence is achieved. If $2k < k_3$, then the integral converges for $\Re(s)<0$ and therefore we should deform the contour $\gamma$ to the left half $s$-plane. On the other hand, if $2k > k_3$, then the integral converges for $\Re(s)>0$ and therefore we should deform the contour $\gamma$ to the right half $s$-plane. Since we are interested in the bispectrum shapes for which  $2k > k_3$, we will deform our original contour $\gamma$ to the right, where $\Re(s)>0$  (see   Fig.~(\ref{fig2})).


Now, let's apply a standard residue analysis to the above contour integral, under the assumption that $2k > k_3$.

Let 
\[
f(s) =   \frac{\cos(\pi s)}{\sin(\pi s)} \frac{\Gamma(1+s)}{\Gamma(\frac{1}{2} + s)}\left(\frac{4k^2}{k_3^2}\right)^{-s}.
\]
Then, according to the \emph{Residue Theorem} from complex analysis, we have that
\[
\frac{1}{2\pi i} \int_{\gamma} f(s)\text{d}s =\frac{1}{2\pi i} \int_{\gamma^+} f(s)\text{d}s = - \sum_{j=1}^{\infty} \text{Res} (f, s_j),
\]
where $s_j$, $j=1,2,3,\dots$ are the poles which are enclosed from the contour $\gamma^+$ and 
\begin{equation}
   \text{Res} (f, s_j) = \lim_{s\to s_j} (s-s_j) f(s).
\label{limit} 
\end{equation}
In our case the poles are known exactly, namely $s_j = j$, where $j=1,2,3,\dots$. Applying L'Hospital's rule in order to compute the limit  \eqref{limit}, one  finds that the contour integral \eqref{contourintegral} is given from an infinite series which can be evaluated exactly. That is, 
\begin{align}
  \mathcal{I}_{iso_1}(k, k_3) &= -\frac{3 \pi^{3/2}}{2 k_3^2}  \times   \frac{1}{2\pi i} \int_{\gamma} f(s) \text{d}s \nonumber\\  &= 
  -\frac{3 \pi^{3/2}}{2 k_3^2}\times  \left[ - \sum_{j=1}^{\infty} \frac{\Gamma(1+j)}{\pi \Gamma(\frac{1}{2}+ j)} \left( \frac{2k}{k_3}\right)^{-2j} \right]\nonumber\\
   &= 
  -\frac{3 \pi^{3/2}}{2 k_3^2} \times \left[- \left( \frac{k_3}{ \pi^{3/2}(4k^2 - k_3^2)} \right)
  \left( k3 + \frac{4k \text{ arccsc}\left(\frac{2k}{k_3}\right)}{\sqrt{4-\frac{k3^2}{k^2}}} \right)
  \right]\nonumber\\
    &= \frac{3 }{2 k_3} \times \left[ \left( \frac{1}{ 
 4k^2 - k_3^2} \right)
  \left( k3 + \frac{4k \text{ arccsc}\left(\frac{2k}{k_3}\right)}{\sqrt{4-\frac{k3^2}{k^2}}} \right)
  \right]. \label{result}
\end{align}
Finally, in the \emph{sqeezed limit}, which corresponds to the local configuration, equation \eqref{result} results to 
\begin{equation}
     \mathcal{I}_{iso_1} = \frac{3}{4 k^2} + \mathcal{O}(k_3),\quad \text{ as } k_3 \to 0, \label{sqeezed}
\end{equation}
while for the \emph{equilateral case} we find
\begin{equation}
      \mathcal{I}_{iso_1} = \frac{9 + 2\sqrt{3} \pi}{18 k^2} + \mathcal{O}(k_3 - k),\quad \text{ as } k_3 \to k. \label{equi}
\end{equation}
The next integral to solve is
\begin{align*}
        \mathcal{I}_{iso_2}(k, k_3) =&
\frac{3  k^2 \pi^2}{8k_3} \int_0^{\infty} \text{d}w\,\, w^2 J_1 (k_3 w) (Y_1(k w))^2 \\
=& \frac{3  k^2 \pi^{3/2}}{k_3^4} \text{MeijerG} \left[\Big\{ \{ -1  \},\{-1/2,1/2 \}\Big\} ,\Big\{ \{ -1,1 \},\{-1/2 \}\Big\}, \frac{4k^2}{k_3^2}
\right],
\end{align*}
which again utilizes the Meijer G-function, as defined in \eqref{generalmeijer}, for $n=1$, $p=3$, $m=2$, $q=3$ and $z= \frac{4k^2}{k_3^2}$, with $a_1=a_n=-1$, $a_2=a_{n+1}=-1/2$, $a_3=a_p=1/2$ and  $b_1=-1$, $b_2=b_{m}=1$, $b_3=b_{m+1}=b_q =-1/2$. Then,
\begin{align}
        \mathcal{I}_{iso_2}(k, k_3) 
 =& \frac{3  k^2 \pi^{3/2}}{k_3^4} \frac{1}{2\pi i} \int_{\gamma} \frac{\Gamma(2-s)\Gamma(s-1)\Gamma(1 +s)}{\Gamma(s-\frac{1}{2}) \Gamma(\frac{1}{2} + s) \Gamma(\frac{3}{2} - s))}\, \left(\frac{4k^2}{k_3^2}\right)^{-s} \text{d}s,\label{withGamma}\\
 =&  \frac{3  k^2 \pi^{3/2}}{k_3^4} \frac{1}{2\pi i} \int_{\gamma} \frac{\cos(\pi s)}{\sin(\pi s)} \frac{\Gamma(1+s)}{\Gamma(\frac{1}{2} + s)}\left(\frac{4k^2}{k_3^2}\right)^{-s} \text{d}s,\label{contourintegral2}
\end{align}
where the contour $\gamma$ is similar to the one illustrated in Fig.~(\ref{fig2}), but shifted to the right of the pole $s_j =1$. As before, for $2k = k_3$ the above integral does not converge, which suggests that either $2k < k_3$ or $2k > k_3$. 

\begin{figure}[h!]
    \begin{subfigure}[b]{0.3\textwidth}
\includegraphics[scale=0.47]{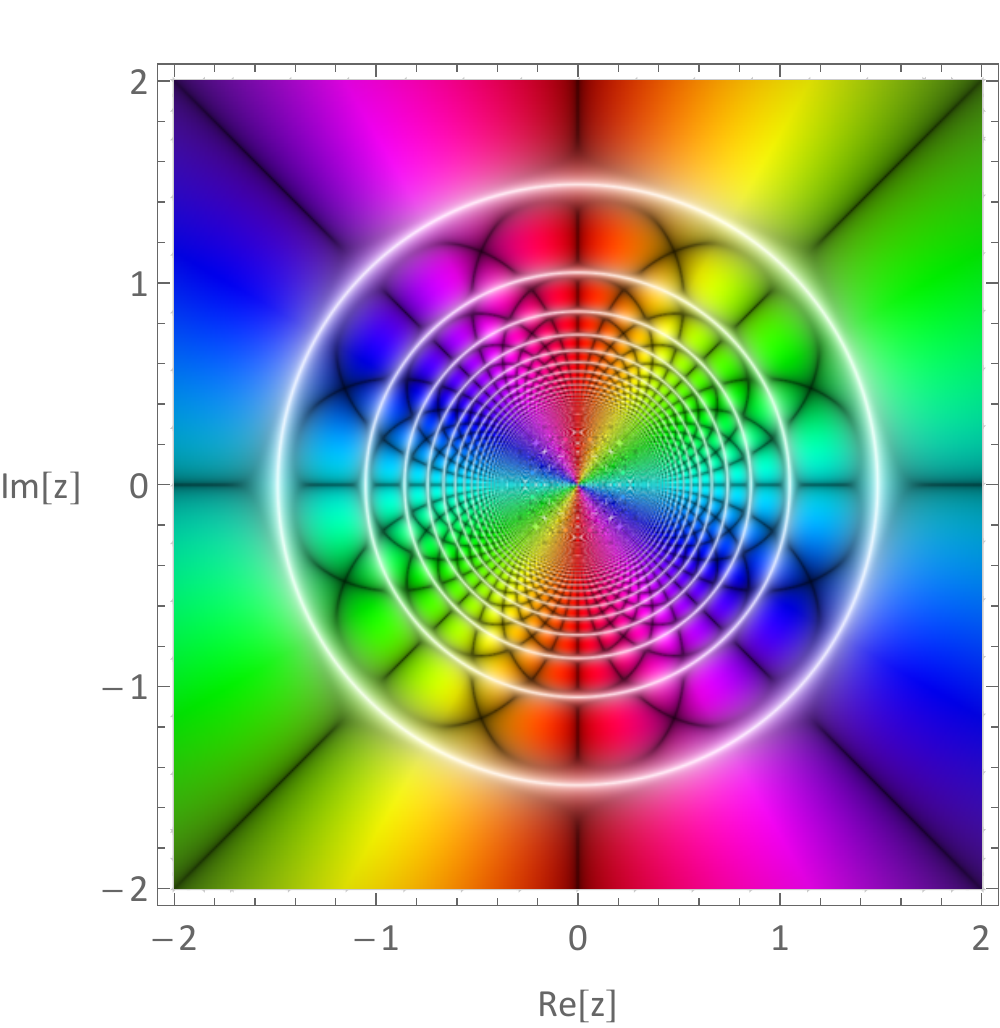}
\caption{}
    \end{subfigure}\hspace{0.7em}
    \begin{subfigure}[b]{0.3\textwidth}
\includegraphics[scale=0.47]{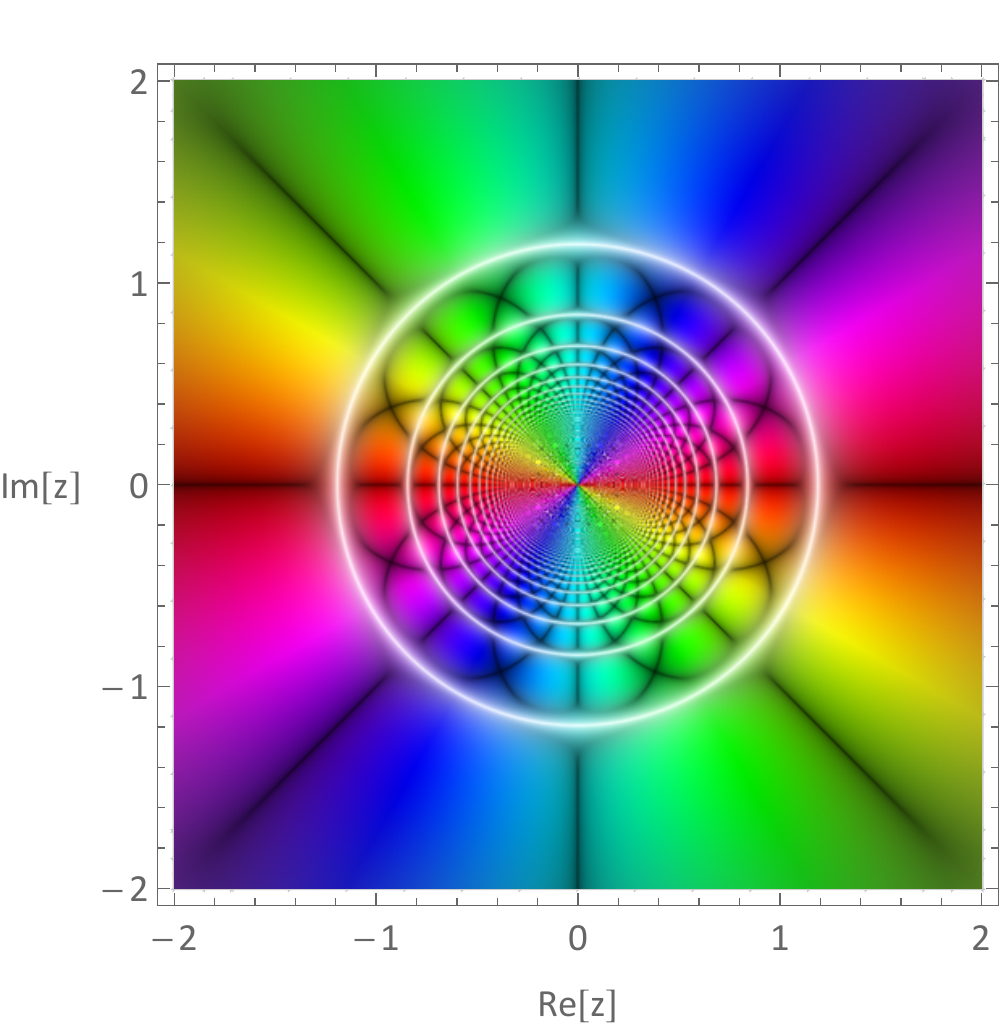}
\caption{}
    \end{subfigure}\hspace{0.7em}
    \begin{subfigure}[b]{0.3\textwidth}
\includegraphics[scale=0.47]{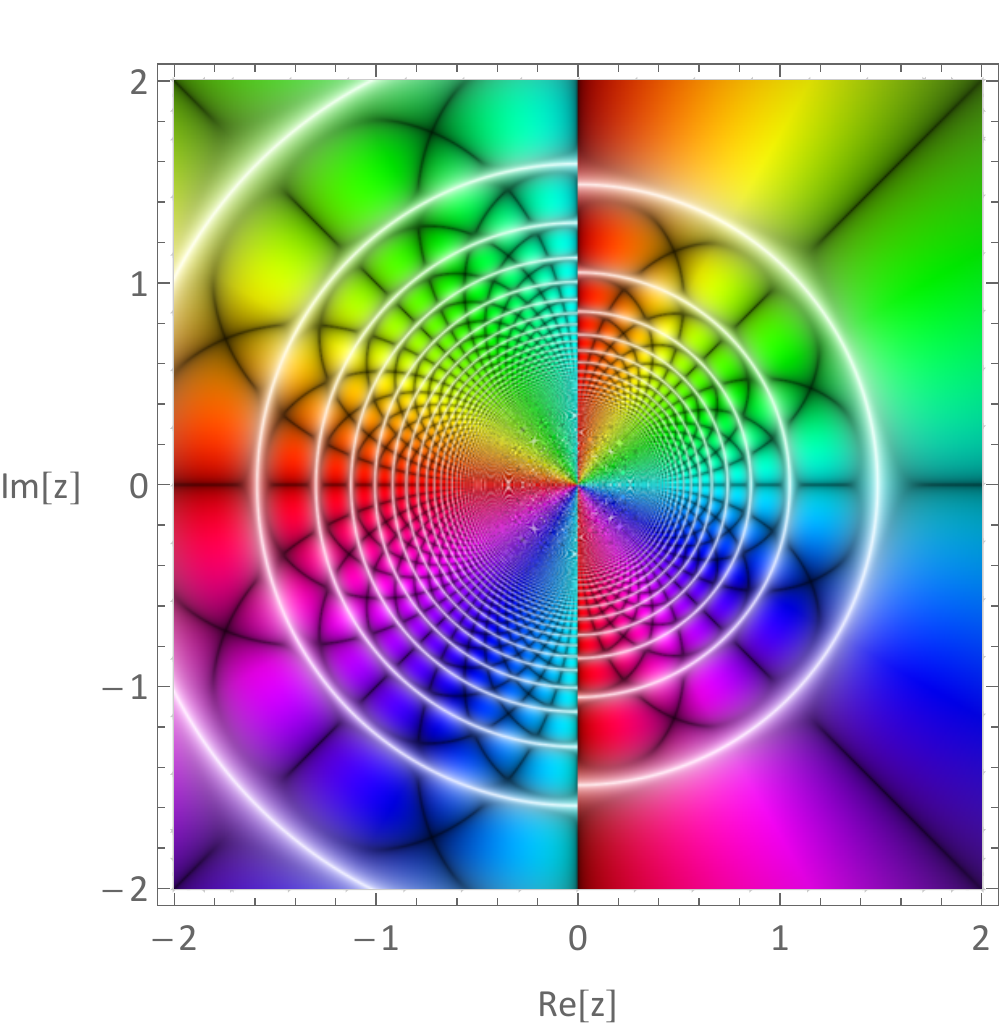}
\caption{}
    \end{subfigure}
    \caption{Phase-portraits of the results  (a) \eqref{result}, (b) \eqref{result2}  and (c) \eqref{result3}. Here, the complex variable is $z=k$. }
    \label{phaseplots}
\end{figure}

Now, let's apply a standard residue analysis to the above contour integral, under the assumption that $2k > k_3$. We will proceed similarly with the analysis of the integral \eqref{contourintegral}. Let
\[
f(s) =   \frac{\cos(\pi s)}{\sin(\pi s)} \frac{\Gamma(1+s)}{\Gamma(\frac{1}{2} + s)}\left(\frac{4k^2}{k_3^2}\right)^{-s}
\]
and apply the \emph{Residue Theorem} 
\[
\frac{1}{2\pi i} \int_{\gamma} = - \sum_{j=1}^{\infty} \text{Res} (f, s_j),
\]
where $s_j$,  ($j=1,2,3,\dots$) are the poles which are enclosed from the contour $\gamma^+$ and 
\begin{equation}
   \text{Res} (f, s_j) = \lim_{s\to s_j} (s-s_j) f(s).
\label{limit2} 
\end{equation}
Likewise with what presented before, the location of the poles is known exactly, namely $s_j = j$, where $j=2,3,4\dots$ \footnote{Note that the first pole to consider is $s_2=2$. This is because of the position of the contour $\gamma$ which is slightly shifted to the right, such that it doesn't include the pole $s_1=1$.}. Applying L'Hospital's rule in order to compute the limit  \eqref{limit2}, one  finds that the contour integral \eqref{contourintegral2} is given from an infinite series which then can be evaluated exactly. That is,
\begin{align}
 \mathcal{I}_{iso_2}(k, k_3) &=  \frac{3  k^2 \pi^{3/2}}{k_3^4}  \times   \frac{1}{2\pi i} \int_{\gamma} f(s) \text{d}s \nonumber\\  &= 
  \frac{3  k^2 \pi^{3/2}}{k_3^4}\times  \left[ - \sum_{j=2}^{\infty} \frac{\Gamma(1+j)}{\pi \Gamma(\frac{1}{2}+ j)} \left( \frac{2k}{k_3}\right)^{-2j} \right]\nonumber\\
   &= 
 \frac{6 k^2 k_3 -3 k_3^3 - \frac{24 k^3 \text{arccsc}\left(\frac{2k}{k_3}\right)}{\sqrt{4-\frac{k_3^2}{k^2}}}}{8 k^2 k_3^3 - 2 k_3^5}.\label{result2}
\end{align}
As we mentioned earlier, all other shapes can be derived from the isosceles configuration by considering the appropriate limits. To retrieve the local configuration we just need to introduce the limiting procedure for $k_3$, while for the equilateral case one should consider $k_3\to k$. One should be careful, since during the limiting process for the equilateral case in the second permutation $I_{iso_2}$, an extra term should be added; a term correcting the complex argument of our result and ensuring that the function maintains its analytical properties. This becomes more clear in Fig.~(\ref{phaseplots}). Let us expand on this argument.

Recall that in the equilateral limit both permutations $I_{iso_1}$ and $I_{iso_2}$ should retrieve the same result. Yet, this is not the case here. Comparing the phase-portraits of results \eqref{result} and \eqref{result2} when $k_3\to k$ ((a) and (b) in Fig.~(\ref{phaseplots})) one observes that the two expressions are not identical, but rather differ in their complex argument. This can be corrected by adding the extra term 
\[
\mathcal{I}_{extra}(k, k_3) = \frac{3\pi k^4}{k_3^3 (4k^2 - k_3^2)^{3/2}}.
\] in \eqref{result2}. Then,
\begin{align}
 \hat{\mathcal{I}}_{iso_2}(k, k_3)&= 
 \frac{6 k^2 k_3 -3 k_3^3 - \frac{24 k^3 \text{arccsc}\left(\frac{2k}{k_3}\right)}{\sqrt{4-\frac{k_3^2}{k^2}}}}{8 k^2 k_3^3 - 2 k_3^5} + 
  \mathcal{I}_{extra}(k, k_3),
\label{result3}
\end{align}
retrieves the correct limit in the right half complex plane, which is made clear by comparing the phase-portraits (a) and (c) in Fig.~(\ref{phaseplots}). Note that we are only interested  in the right half--plane, which is the domain where $\Re(k)>0$.

Finally, in the limiting cases of interest one obtains
\begin{equation}
     \mathcal{I}_{iso_2} = -\frac{1}{2 k^2} + \mathcal{O}(k_3),\quad \text{ as } k_3 \to 0, 
\end{equation}
and
\begin{equation}
      \mathcal{I}_{iso_2} = \frac{9 + 2\sqrt{3} \pi}{18 k^2} + \mathcal{O}(k_3 - k),\quad \text{ as } k_3 \to k, 
\end{equation} where in the last result we have dropped the hat notation.

\subsection{Regulating oscillatory integrals}
		\label{Ap2}

\begin{figure}
	\begin{subfigure}[b]{0.45\textwidth}
		\centering
		\includegraphics[width=\linewidth]{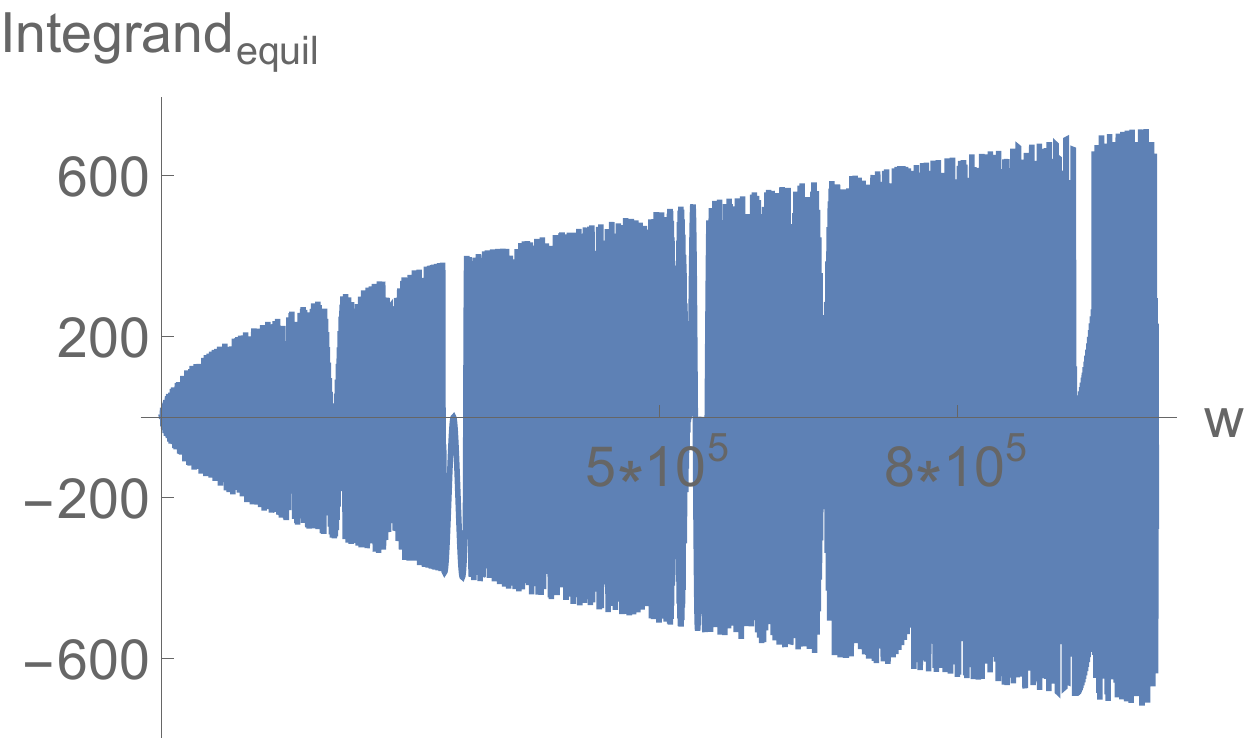}
		\caption{ Without regulator}
		\label{fig:equra}
	\end{subfigure}\hfill
	\begin{subfigure}[b]{0.45\textwidth}
	\centering
	\includegraphics[width=\linewidth]{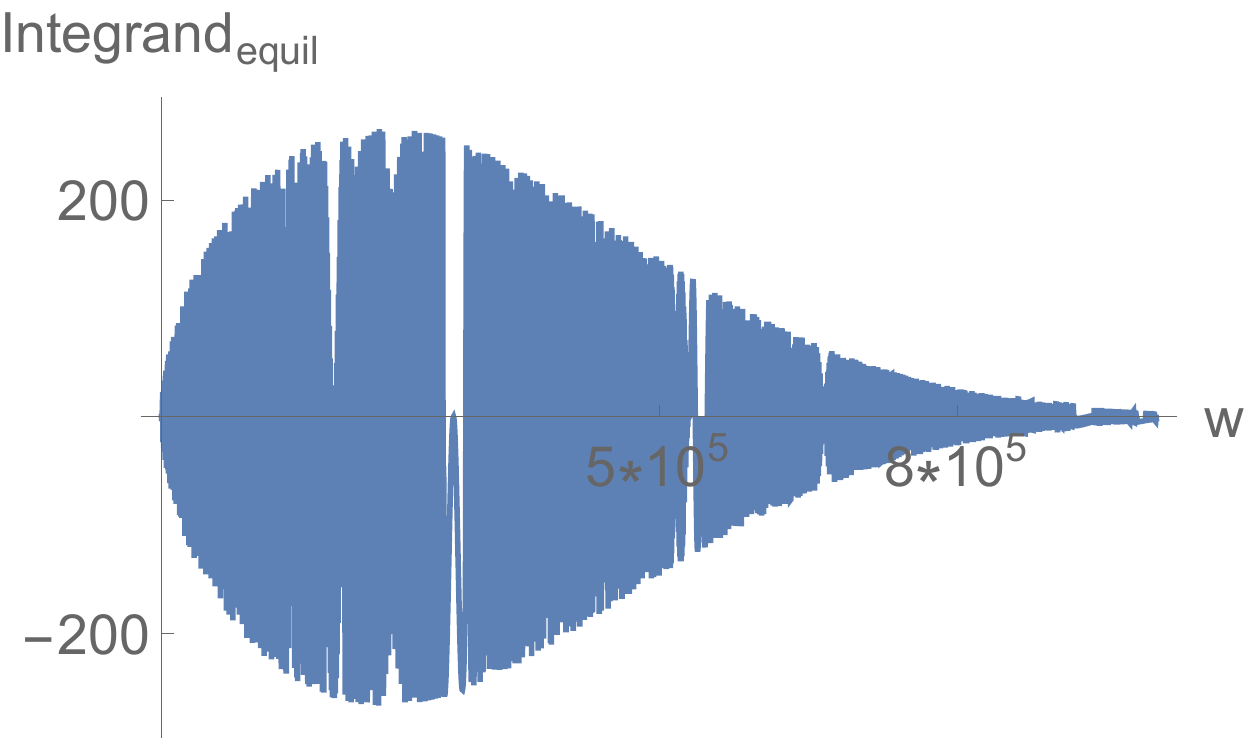}
	\caption{With regulator}
	\label{fig:equrb}
\end{subfigure}
	\caption{The integrand in the equilateral configuration can be regulated at infinity. In Fig. (a) we plot the integrand without a regulator, while in Fig. (b) we plot it with a regulator.} \label{fig:equr}
\end{figure}

		The integrals in (\ref{eq:i8}) are highly oscillatory and suffer from convergence issues. A \textit{regulator}, or damping factor, can force decaying behaviour  which allows us to compute the integral numerically. Take, for example, the integral in the equilateral configuration, given in (\ref{eq:i9}). We multiply the integrand by a term of the form $e^{- \epsilon \abs{w}^2/2}$, where $\epsilon \ll 1$
		\begin{equation}\begin{split} 
\mathcal{I}_{equil} &=  \frac{3 k \pi^2}{8} \int_0^{\infty}  \dd{w}  w^2  J_1(k w) Y_1^2(k w)  e^{-  \frac{\epsilon\abs{w}^2}{2}}.
\label{eq:app1} 
\end{split}\end{equation}
As seen in Fig. (\ref{fig:equra}), the integrand is highly oscillatory with increasingly large amplitudes as $w$ approaches infinity. Using the regulator, forces the integrand to decay at infinity as seen in Fig. (\ref{fig:equrb}). This helps the integral to converge at infinity and therefore, to numerically compute it, giving full agreement with the analytic result. 

\begin{figure}
	\begin{subfigure}[b]{0.45\textwidth}
		\centering
		\includegraphics[width=\linewidth]{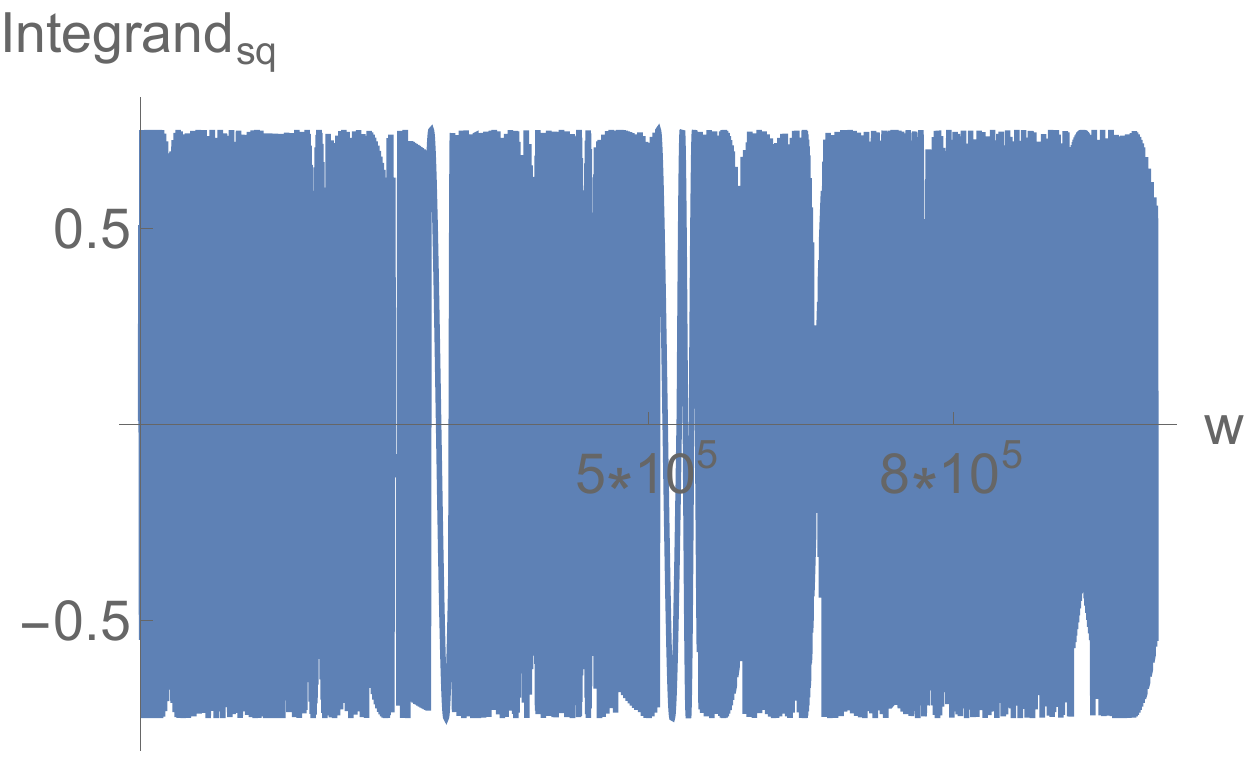}
		\caption{ Without regulator}
		\label{fig:sqra}
	\end{subfigure}\hfill
	\begin{subfigure}[b]{0.45\textwidth}
	\centering
	\includegraphics[width=\linewidth]{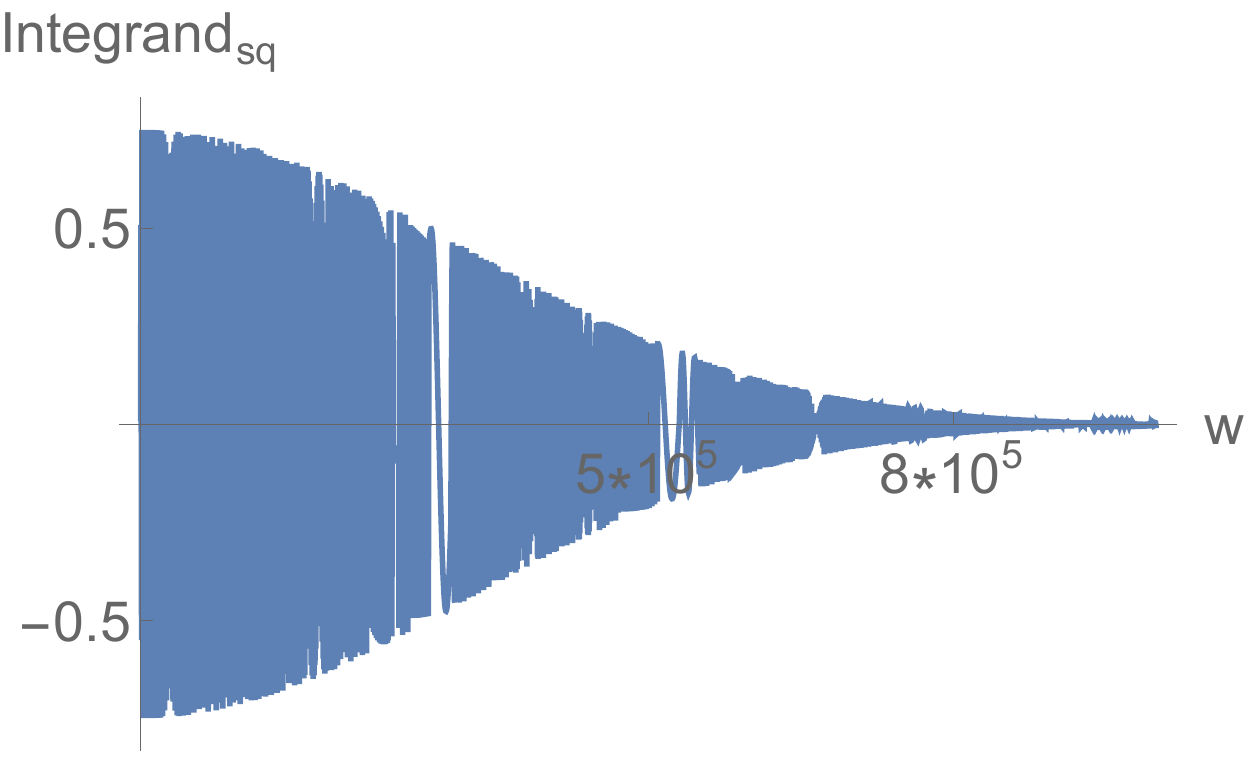}
	\caption{With regulator}
	\label{fig:sqrb}
\end{subfigure}
	\caption{The integrand in the squeezed configuration, for the first two permutations, can be regulated at infinity. In Fig. (a) we plot the integrand without a regulator, while in Fig. (b) we plot it with a regulator.} \label{fig:sqr}
\end{figure}

Similarly, the oscillatory behaviour of the integral in the squeezed limit, given in (\ref{eq:i11}), can be regulated at infinity, using the same method, as seen in Figure (\ref{fig:sqr}). This ensures convergence for $w \rightarrow \infty$ in the numerical computation.

The numerical result becomes more accurate as $\epsilon$ takes smaller  values and it is valid for a large range of comoving scales seen by CMB experiments.

\begin{figure}
	\begin{subfigure}[b]{0.45\textwidth}
		\centering
		\includegraphics[width=\linewidth]{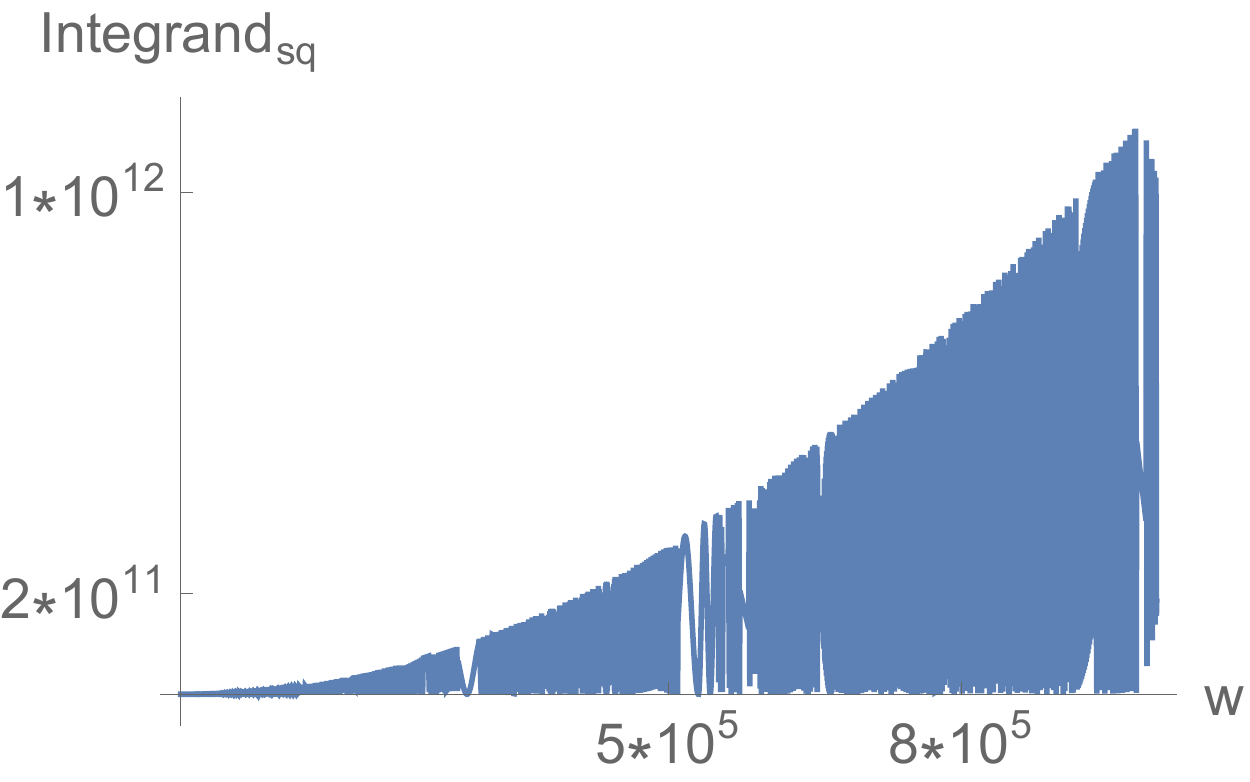}
		\caption{ Without cut-off}
		\label{fig:sqra2}
	\end{subfigure}\hfill
	\begin{subfigure}[b]{0.45\textwidth}
	\centering
	\includegraphics[width=\linewidth]{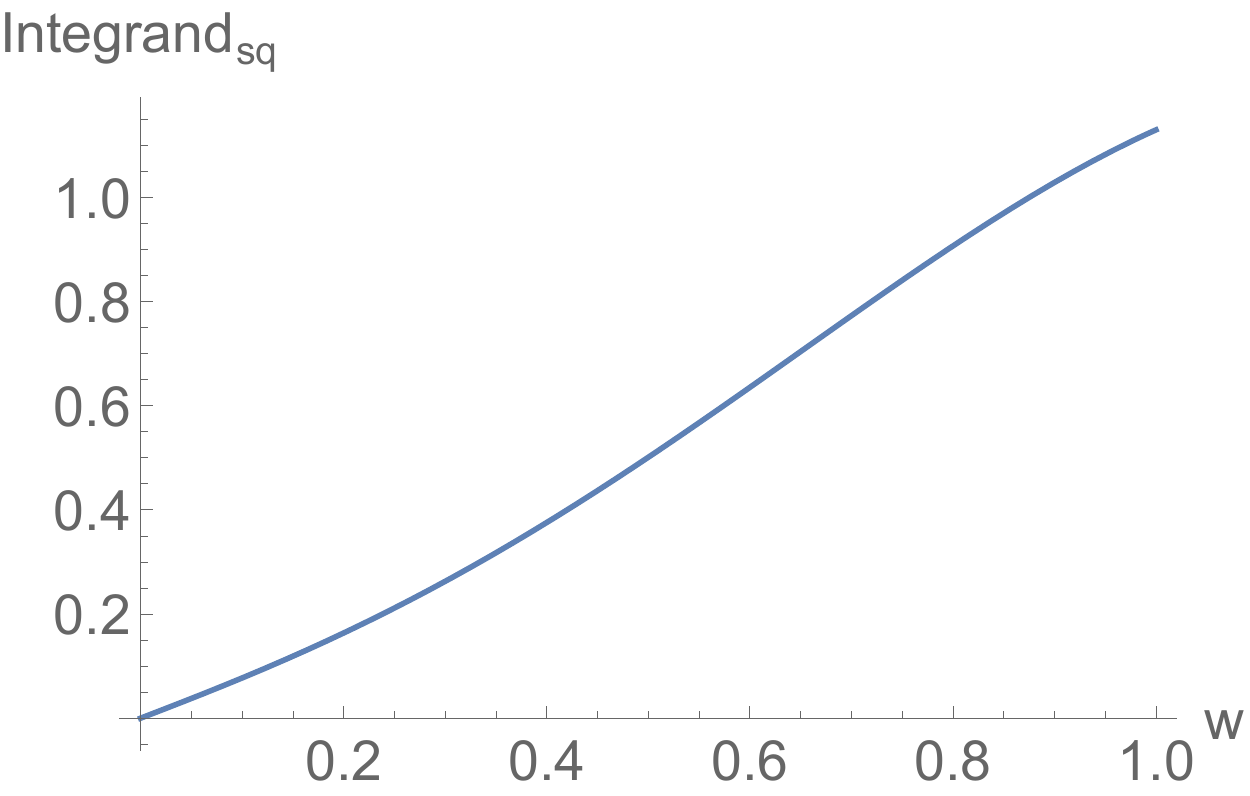}
	\caption{With cut-off}
	\label{fig:sqrb2}
\end{subfigure}
	\caption{The integrand in the squeezed configuration, for the last permutation, can be regulated at infinity by using a hard cutoff. In Fig. (a) we plot the integrand without a hard cutoff, while in Fig. (b) we plot it with a hard cutoff.} \label{fig:sqr2}
\end{figure}

Unfortunately, the regulator is not effective on all cases considered here. Instead, we can use a hard cut-off $w_\Lambda \sim 1/k$. (see Fig. (\ref{fig:sqr2})). For suitable choice of $w_\Lambda$, this successfully approximates the analytic result at $w \rightarrow 0$ in (\ref{eq:i18}), for a wide range of modes.


	\bibliographystyle{JHEP}
	\bibliography{bib.bib}
\end{document}